%% file: something.tex
\input some_psfig

\input some_incl

\newdimen\subskipamount  \subskipamount=10pt

\fignumber=1
\tabnumber=1
\eqnumber=1

\def\RADIO{11}
\def\BDV{6}

\def\SA{S97A} % Sevenster \etal1997a}
\def\SB{S97B} % Sevenster \etal1997b}
\def\SDH{SDH} % Sevenster \etal1995}
\def\CHTWO{Sevenster \etal1997a,b}
\def\CHTHR{Sevenster \etal1999} 
\def\CHFOU{Sevenster 1997}
\def\CHFIV{Sevenster \etal1995}

\def\artp{article}
\def\artc{article}

\def\RRmi{R$_{-1}$}
\def\RRpi{R$_{+1}$}
\def\RRmu{R$_{-2}$}
\def\RRpu{R$_{+2}$}
\def\RRmc{R$_{-3}$}
\def\RRpc{R$_{+3}$}
\def\RRmp{R$_{-4}$}
\def\RRpp{R$_{+4}$}
\def\vs{\char'24\hskip -4pt {s}\ }
\def\Pal{{Palou\char'24\hskip -4.3pt {s}}\ }

\def\edott{$\epsilon_{\rm \dot M}$ }

\def\azh{apparent scaleheight}
\def\zh{scaleheight}
\def\arh{apparent scalelength}
\def\rh{scalelength}

\def\S{Sect.}

\input mn
\BeginOpening

\title{Something about the structure of the Galaxy}

\author{Maartje N. Sevenster$^{1,2}$} % \authorcomment{2 Now at RSAA/MSSSO}}

\vskip .5truecm
\affiliation{$^1$Sterrewacht Leiden,
POBox 9513, 2300 RA Leiden, The Netherlands}
\vskip .1truecm
\affiliation{$^2$Presently at : RSAA/MSSSO, Private Bag Weston Creek PO, Weston 2611 ACT,
Australia (msevenst@mso.anu.edu.au)}

\shortauthor{M.~Sevenster}
\shorttitle{Something about the Galaxy}

\acceptedline{Accepted xx. Received xx}

\abstract{
We analyse a sample of 507
evolved (OH/IR) stars in the region 
($ 10^{\circ} > \ell > -45^{\circ} $),($ |b| < 3^{\circ} $). 
We derive average ages for
subsets of this sample and use those sets as
beacons for the evolution of the Galaxy.
In the Bulge the oldest OH/IR stars in the plane
are 7.5 Gyr (1.3 \msun), in the Disk 2.7 Gyr (2.3 \msun).
The vertical distribution of almost all AGB stars
in the Disk is found to be nearly exponential, with
scaleheight increasing from 100 pc for stars of 
\lsim 1 Gyr to 500 pc for stars of \gsim 5 Gyr. 
There may be a small, disjunct population of OH/IR stars.
The radial distribution of AGB stars is dictated by 
the metallicity gradient.
Unequivocal morphological evidence is presented for the existence of a 
central Bar, but parameters
can be constrained only for a given spatial--density model.
Using a variety of indicators, we identify
the radii of the inner ultra--harmonic (2.5 kpc) and corotation
resonance (3.5 kpc).
We show that the 3--kpc arm is likely to be an inner ring, 
as observed in other barred galaxies,
by identifying a group of evolved stars that is connected to the
3--kpc HI filament.
Also, using several observed features, we argue that an 
inner--Lindblad resonance exists, at $\sim$1--1.5 kpc.
The compositions of OH/IR populations within 1 kpc from the galactic Centre
give insight into the bar--driven evolution of the inner regions.
We suggest that the Bar is $\sim$8 Gyr old, relatively weak (SAB)
and may be in a final stage of its existence. 
}

\keywords{Galaxy: structure -- Galaxy: evolution -- Galaxy: stellar content
   -- Stars: AGB and post--AGB.}

\maketitle

\section{Introduction}

It has become widely accepted that our Galaxy is barred,
as evidence accumulated over the last five years 
from star counts, gas--dynamical studies and stellar three--dimensional
kinematics and especially from the analysis of the COBE--DIRBE
integrated--light data (Dwek \etal1995; for a review on the
\gba\ see Gerhard 1996). 
The important parameters of a barred potential are the 
semi--major--axis length, $a$, the in--plane 
axis ratio, $q$, the pattern speed, \pspeed, and the strength 
relative to the axisymmetric part of the potential, $A$. 
From the observer's point of view, another quantity is
the major axis' orientation with respect to the \losn , the 
viewing angle $\phi$.  Due to the awkward view we have
of the Galaxy, the values of those parameters are even
harder to determine than in external galaxies. They are often
inferred from the influence the barred potential has on the
other parts of the galaxy. Helpful, albeit crude and not completely
understood, diagnostics are the ``resonant rings''
and spiral features
that arise at radii where orbits are in resonance with
the frequency of rotation of the bar. The use of these structures
is especially difficult in our Galaxy, where we see only 
tangent points to rings and spiral arms, but we
will attempt to do so.

The aim of this article is to describe
the overall form of the stellar distribution in, and the evolution
of various galactic components (disk, bulge, spiral arms etc), based 
mainly on its content of evolved stars and focussing mainly on the
inner Galaxy. To a lesser
extent, we attempt to constrain the free parameters of these stellar
distributions. The goal is a schematic, rather than comprehensive,
picture of the Galaxy.

We restrict ourselves largely to the stellar distribution
and discuss the distribution of the gas only superficially.
As stars are probably the source and the cleanest tracers of the 
barred potential, such a study of only the stellar component 
of the Galaxy is necessary.
Obviously, the gas-- and the
stellar distributions should ultimately be explained simultaneously
in one coherent picture.
In fact, the first evidence for non--axisymmetry of the inner Galaxy
came from the neutral--gas kinematics (see review by Oort 1977). It was 
exactly the drive for coherence that has shifted the attention
to the stars, because their distribution 
showed no clear deviation from axisymmetry.
Neither of the prevailing explanations for the origin of
the observed radial gas
motions - central expansion or elongation of the potential - were supported by 
stellar observations.
Once the Bar had been found in the stellar surface 
density (Blitz \& Spergel 1991; Dwek \etal1995),
questions remained. For instance, why are the stellar kinematics 
explained so well by axisymmetric models (Kent 1992;
Ibata \& Gilmore 1995) and why is the micro--lensing
optical depth toward Baade's window incompatible
with density models derived from surface--density maps (eg. Nikolaev
\& Weinberg 1997) ?

The stellar data used in this paper (\CHTWO) were collected 
specifically to form the optimal sample to address such issues
and to complement existing data. 
The sample consists of OH/IR stars: intermediate--mass, 
oxygen--rich, far--evolved asymptotic--giant--branch (AGB) stars.
These are excellent tracers of the general stellar population,
as stars with initial masses between roughly 1\msol\ and 6\msol\ go
through this phase. They are also tracers of the galactic potential
as they form a fairly relaxed population, with typical ages of several gigayears.
Very strong and characteristic
maser emission at 1612 MHz from the ground--state
OH molecule allows for radio--interferometric observations, 
combining extinction--free coverage of the plane and fast sampling out to large
distances with positions and velocities
with negligible errors (compared to modelling errors).
The employability of this sample in galactic--structure studies
is clearly demonstrated in \CHTHR .

The structure of this paper is as follows.
In the first part, we discuss the large--scale spatial distribution
of the OH/IR--star sample.
We describe the sample of OH/IR stars in more detail in \S 2.
In \S 3 we analyse the structure of this sample and estimate
the scaleheights and --lengths for the \gbu\ and \gd. 
Variations of those exponential scales with position are combined with scales
derived from similar samples and interpreted as age dependencies.
In \S 4, we give morphological evidence for a
\gba\ that cannot be 
explained by a physical lopsided distribution (see Blitz \& Spergel 1991; Sevenster 1996).
By comparing some parametrized models to the observations,
a flavour of the values of the relevant parameters is obtained. Most importantly, we
derive an approximate distance limit for the sample.
The kinematic type of the \gbu\ 
in terms of its anisotropy parameter and ellipticity
is found.

In the second part of the paper we discuss the influence of the bar--shaped
potential on the inner Galaxy.
In \S 5 we interpret various patterns found
in the sample as resonant patterns and identify resonant radii.
New insight into the nature of the so--called 3--kpc arm is provided
by a small group of OH/IR stars that follows the 3--kpc arm's 
longitude--latitude--velocity structure.
In \S 6 we give a description of the inner Galaxy based
on the presented evidence and we speculate on
its evolution in \S 7. We conclude in \S 8.

Throughout this \artp, we will use the term ``Bulge'' to 
denote the galactic component we see in the general direction 
of $ |\ell|$ \lsim 10\degr, without being interested in its actual form.
If we use the term ``Bar'' we denote specifically the 
prolate or triaxial component of the Bulge.
Unless explicitly stated otherwise,
stellar velocities are given with respect to the local standard
of rest (see \CHTWO\ for the used Doppler corrections) and $R_{\odot}\equiv$8 kpc.

\section{Properties of the data set}

\beginfigure*1
\fignam\SURF
%%%{\psfig{figure=fivesurface.ps,height=0.9\hsize,angle=270}}
\hskip .5truecm{
\psfig{figure=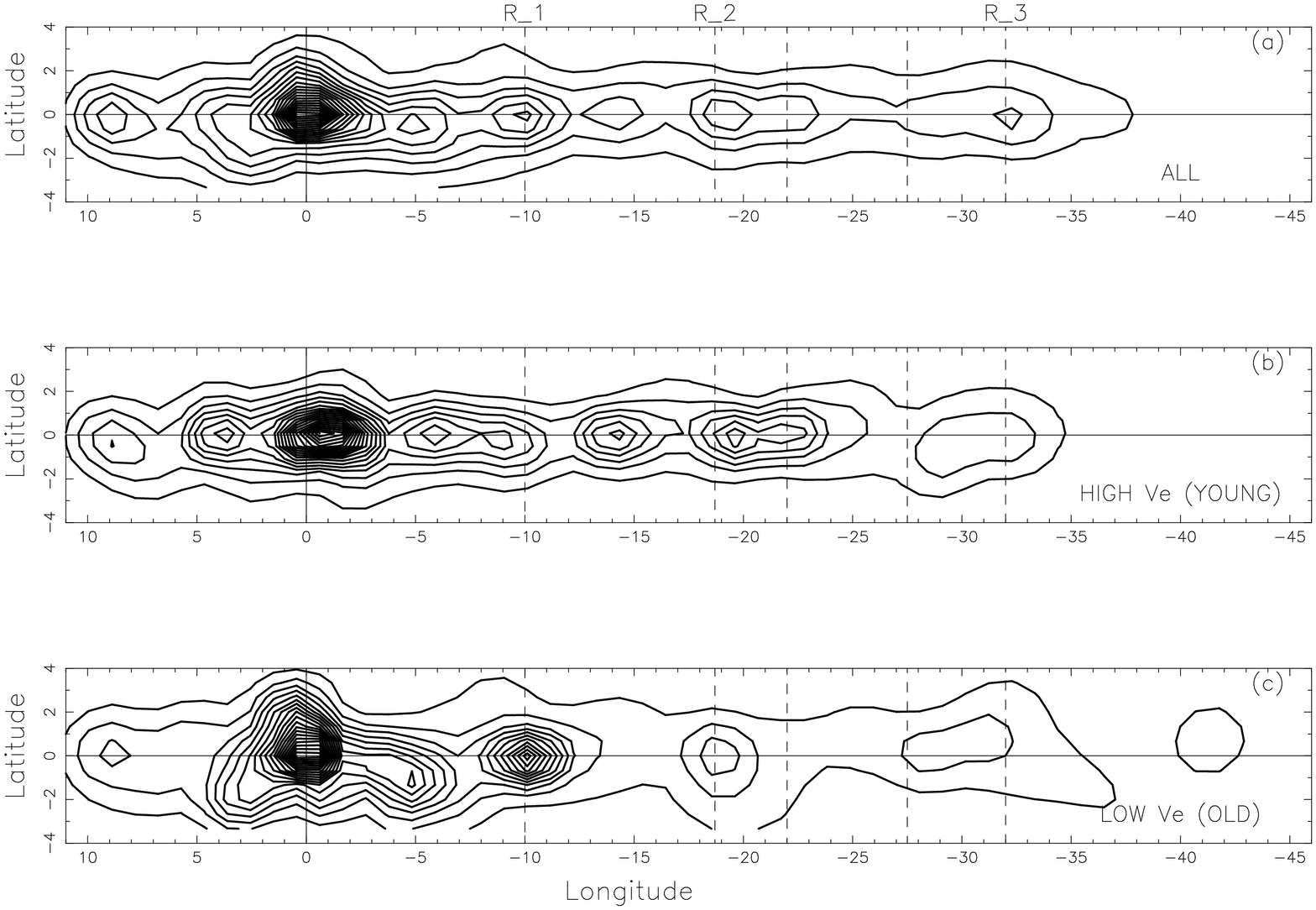,height=0.65\hsize,angle=270}}
\caption{{\bf Figure \nfig}
The surface density of the AOSP sample (\S 2), in
longitude versus latitude in degrees. The observations
were smoothed with an adaptive--kernel algorithm (initial kernel of 1\deg ). 
The upper panel is for all OH/IR stars (507), 
the middle for outflow velocities
higher than 14 \kms\ and the lowest panel
for outflow velocities lower than 14 \kms (and higher than 1 \kms).
Panels {\bf b, c} represent younger and older stars, respectively,
with an abundance effect (\S 2).
Both subset--plots are based on $\sim$210 stars.
Note the offset toward negative longitudes in for the young sample (see \S 4.1).
The local maxima at $\sim -10^{\circ},-20^{\circ},-30^{\circ}$ --
\RRmi, \RRmu\ and \RRmc, respectively -- will be discussed in \S 5. 
The unmarked vertical dashed lines indicate maxima in the 
observed 2.4 GHz distribution (see \S 5, Fig.\RADIO).
The maximum at $\ell = +10^{\circ}$ is not reliable.
Contours are spaced at twenty even intervals
between zero and the maximum of each particular plot.
}
\endfigure

\noindent
The stellar density and the gravitational
potential of the Galaxy are best traced by intermediate--mass,
evolved stars as they constitute the largest fraction of 
the total stellar mass and are dynamically relaxed (Frogel 1988). 
Good candidates for this are
the so--called OH/IR stars;
oxygen--rich objects on the asymptotic giant branch (AGB;
see for instance Habing 1993; Sevenster, Dejonghe \& Habing 1995).
We use an unbiased, homogeneous sample of 507 OH/IR stars in the galactic 
Plane, the AOSP sample (Australia telescope Ohir Survey of the Plane),
acquired in a systematic survey of
the region between longitudes $ 10^{\circ} > \ell > -45^{\circ}$ and
latitudes $ |b| < 3^{\circ}$
in the 1612 MHz (18 cm) OH--maser line (\CHTWO).
The positional accuracy is \decsec0.5, the \losa\ velocities (with
respect to the local standard of rest) are accurate to 1 \kms.

In Fig.\SURF (a) we show the surface density
of the AOSP sample, smoothed with an adaptive--kernel algorithm 
(Merritt \& Tremblay 1994). 
With the initial kernel size that retains best
the steepness of the central density
profile without showing individual stars
(1\deg) many local maxima are revealed in the
distribution (\RRmi , \RRmu \ and \RRmc).
The maximum at $\ell = +9^{\circ}$  is likely to be a spurious
edge effect.
We will not discuss any asymmetries in the vertical direction, such
as a tilt, because the data quality is slightly 
latitude--dependent (\CHTWO).
First, in this section, 
we will assess some astrophysical properties of the sample.

\beginfigure2
\fignam\BCUM
{\psfig{figure=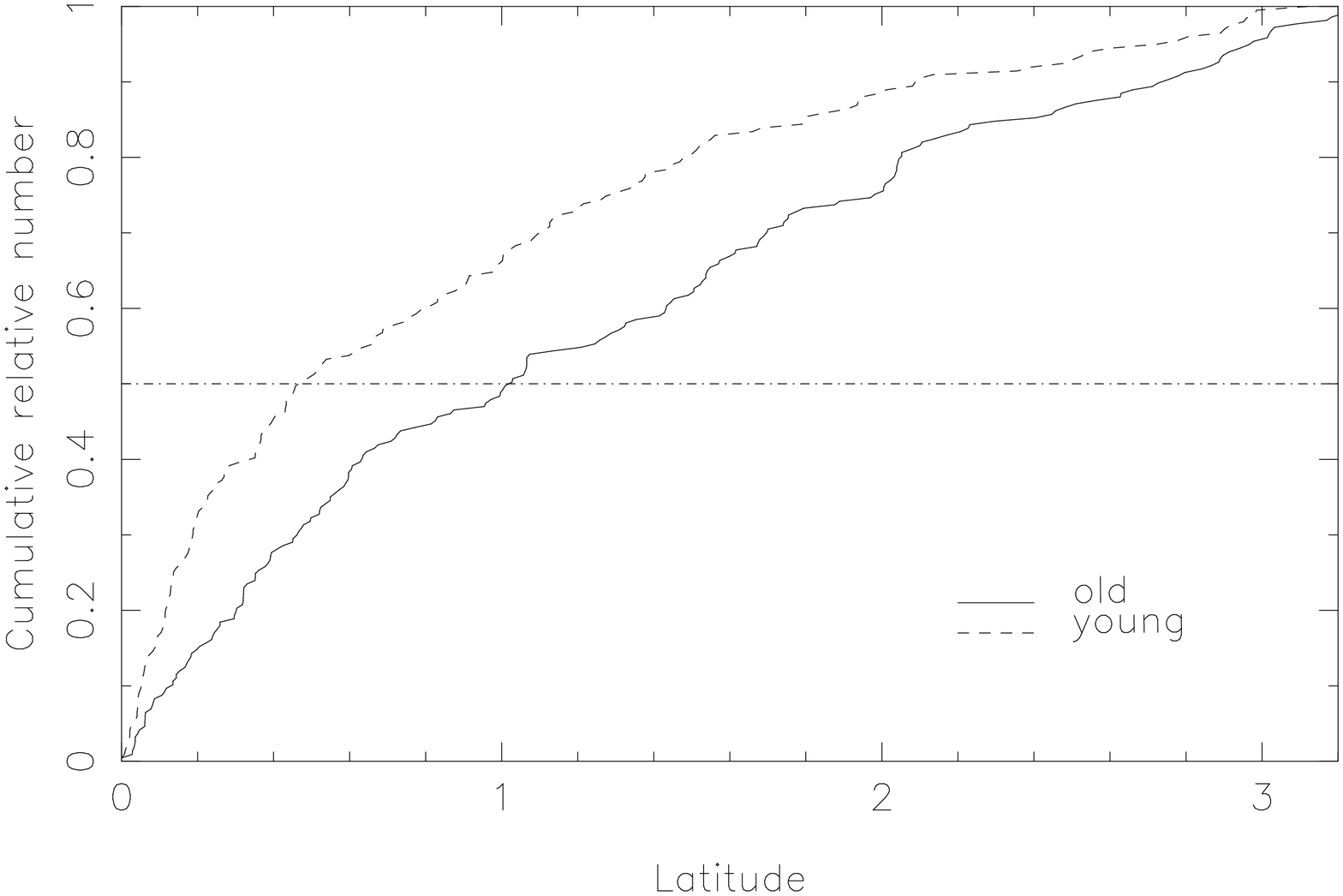,width=\hsize,angle=270}}
\caption{{\bf Figure \nfig}
The cumulative--number distribution over latitude for the two
subsamples of the AOSP sample (Fig.\SURF (b,c)).
The (apparent) scaleheights differ by a factor of 2, which is
indicative of the age difference between the samples.
}
\endfigure

The OH emission comes from an optically--thick, expanding \cse .
Because the expansion is radiation--pressure driven,
the unobservable intrinsic stellar luminosity $L_{\ast}$ is related
to the outflow velocity of the \cse , \vexp , and 
its gas--to--dust ratio, $\mu$, (related to
metal abundance $Z$; $\mu \propto Z^{-1}$ for oxygen--rich stars, 
Habing, Tignon \& Tielens 1994). According to van der Veen (1989):
\eqnam\MU$$
  L_{\ast} \propto \mu^2 V_{\rm exp}^{4} .
%%%%  L_{\ast} \propto \mu^{1.7} v_{\rm exp}^{3.3}  (HTT)
\eqno(\new) $$
This equation is derived and discussed in more detail in Appendix A.
Separating the objects according to outflow velocity (Fig.\SURF (b,c))
hence results in a separation roughly according to stellar luminosity, or age,
without knowledge of the distances to the objects.
This results, as expected, in very different
apparent scaleheights of the subsamples (Fig.\BCUM , see discussion in \S 3).

OH/IR stars can span a wide range of ages, of 0.1 Gyr to \gsim 10 Gyr.
From the IRAS two--colour diagram, 
one can determine a ``turn--over'' [25]--[60] colour $R^f_{32}$,
of sources leaving the evolutionary track (van der Veen \& Habing 1988).
This $R^f_{32}$ is related to the initial mass (Garcia Lario 1991);
this relation is discussed in Appendix B.
We can thus find the initial mass of a star that has reached the
end of the OH/IR--star phase (the thermally--pulsing AGB phase).
Since this phase itself is short ($\sim10^5$yr, Tanabe \etal1997) 
compared to
the ages of the AGB stars, we can use the value for the AGB--tip ages 
by Bertelli \etal(1994) to obtain, for an assumed abundance,
the age of the star from its initial mass.
The minimum $R^f_{32}$ for a sample hence gives the maximum
age of the stars in that sample. We determined this for the Bulge region 
($R^f_{32}>-$0.4, \CHTWO) and the Disk region ($R^f_{32}>+$0.2, \CHTWO)
separately. We find minimum initial masses and maximum ages of 1.3\msun \&
7.5 Gyr and 2.3\msun \& 2.7 Gyr, respectively, both for solar abundance.
The ratio of the mean initial masses (3.3/2.3 with an IMF $\propto M^{-2.5}$,
almost independently of $Z$ used to derive the ages)
and of mean outflow velocities (15/14 \kms) are compatible with a similar
average abundance for Disk
and Bulge stars, $Z_{\rm b} =$ 0.9--0.95$ Z_{\rm d}$, according to 
equation (\MU).
The number of stars in the AOSP sample with ages \lsim 0.5 Gyr 
is negligible.

These values are meant to give an indication only, with a likely uncertainty
of about 15\% in the derived ages and about 10\% in the initial masses.
The $R^f_{32} - M_i $ relation is not yet
fully established and should be used only for ensembles, not
for individual objects. 
Besides, not all AOSP stars have a (reliable) IRAS 
identification (\CHTWO), although those that do 
must be representative of those 
that don't, as the identifications are hampered mainly
by the confusion--limited spatial resolution of IRAS.

The ages are influenced by the fact that the AOSP sample covers
only low latitudes. The fraction of (sub--) solar--mass OH/IR stars
is apparently smaller than 1\% at low latitudes. Out of the plane,
van der Veen \& Habing (1990) find the masses of oxygen--rich
AGB stars to range to $<$ 1.0 \msun, and the corresponding ages
to well over 10 Gyr. However, with   
their mass-loss parameter {\edott=1.5} instead 
of {\edott=1.0} (both values are 
equally likely to be right), the mass range would be
1.2--2.2 \msun, which fits in very well with our derivations.
Correspondingly, the ages of their stars 
would be lower, especially when using $Z=0.02$
instead of $Z=0.04$ as they advocate. 
We conclude that our results are compatible with theirs.

\section{Exponential scales of the Disk and the Bulge}

\beginfigure3
\fignam\SCAL
{\psfig{figure=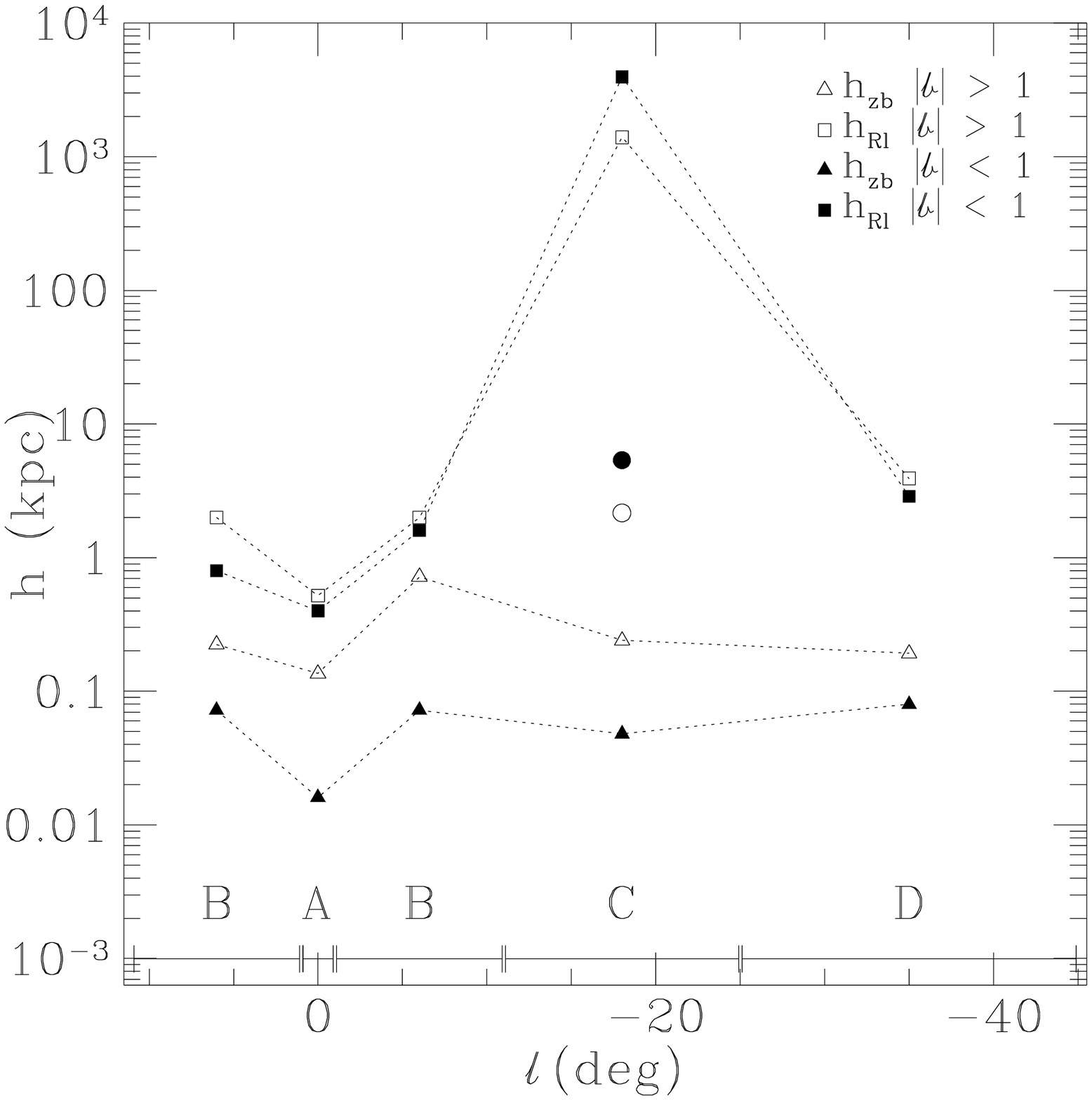,height=8cm}}
\caption{{\bf Figure \nfig}
The local apparent exponential scaleheights 
$h_{{\rm z}b}$ and $h_{\rm R\ell}$
of the AOSP sample, from fits to the cumulative distributions in 
latitude and longitude separately.
The symbols are 
for $ 0 < |b| < 1$ (closed) and $1 < |b| < 3$ (open), respectively.
The bars along the abscissa indicate the longitude
bins (A,B,C,D) used for the determination of the fits; the data points are at
the middle values of these bins. All longitude
bins contain about 100 stars
over the full latitude range ($|b| < 3^{\circ}$); the formal
errors on the measurements are $\sim$15\%.
The two circles are the scalelengths fitted to low--outflow stars
(cf. Fig.\SURF (c)) only. 
For A the apparent scaleheight is very small due to the presence
of the galactic--centre population (see \S 3.2). 
Very large apparent scalelengths, such as in longitude bin C,
are equivalent to an infinite scalelength or a (locally)
flat distribution (\S 3.1).
}
\endfigure

\noindent
The latitude distribution of the AOSP sample is steeper
than a (projected) exponential (${\rm exp}(-z)$) at
all longitudes and irreconcilable with
flatter functional forms (${\rm sech^2}(z)$ or ${\rm exp}(-z^2)$).
The same was found recently by de Grijs \& Peletier (1997) 
for a large number of spiral galaxies; Kent, Dame \& Fazio (1991)
found from NIR observations of
the Galaxy that the profile is closer to exponential than
to ${\rm sech^2}(z)$.
To obtain estimates of the values of the 
scalelength and scaleheight of the spatial density, 
we hence used double exponentials 
($\rho \propto {\rm exp}(-R/h_{\rm R})\,{\rm exp}(-z/h_{\rm z})$).
We estimated those values in various regions of the sky, to
assess variations of the scales with radius and height (age).
Note, however, that we are not trying to model the Galaxy as a
set of double exponentials.

We fitted cumulative--number densities 
with single exponentials, in latitude (in ${\rm R_{\odot}}\tan b$);
${\rm R_{\odot}} \equiv$ 8 kpc) 
and longitude (in ${\rm R_{\odot}}\sin \ell$) separately.
This yielded apparent scales $h_{{\rm z}b}$
and $h_{\rm R \ell}$ of the local surface--density distribution at
various longitudes and latitudes (Fig.\SCAL).
Due to the non--trivial angles between lines of sight,
the scales are not invariant for projection,
even though the Galaxy is seen edge--on.
This means we have to deproject the apparent scales to 
find the true scales $h_{\rm z}$ and $h_{\rm R}$.
By projecting analytic double--exponential distributions
with a range of $h_{\rm R}$ and $h_{\rm z}$, we obtained a range
of apparent scales at the same ($\ell,b$) as the data measurements.
We could thus retrieve the intrinsic ($h_{\rm z}$, $h_{\rm R}$)
that would yield a measured pair ($h_{{\rm z}b}$, $h_{\rm R \ell}$).
Projection effects cause the relation between the true and 
the apparent scales to depend
upon longitude and latitude, and also upon each other, thus
jeopardizing a unique deprojection.

\begintable*1
\tabnam\HRZ
\caption{{\bf Table \ntab.} 
Deprojected exponential 
scales for the squares and triangles in 
Fig.\SCAL\ (columns 3--6) and for 
low-- (``o'', columns 7,8) and high--outflow (``y'', columns 9,10)
sources ($|b| < 3^{\circ}$).
The values between brackets are from another sample at 
$|\ell| = 0^{\circ}$ (see text) or conceivably unreliable because of the
infinite value of $h_{\rm R}$ at $|\ell|=18^{\circ}$.
}
\tabskip=1em plus 2em minus 0.5em%
\halign to 15cm{
\hfil$#$\hfil&\hfil$#$\hfil&\hfil$#$\hfil&\hfil$#$\hfil&\hfil$#$\hfil&
  \hfil$#$\hfil&\hfil$#$\hfil&\hfil$#$\hfil&\hfil$#$\hfil&\hfil$#$ \cr
\noalign{\vskip2pt\hrule\vskip2pt\hrule\vskip2pt}
 & |\ell| & h_{\rm R} (|b|<1^{\circ}) & h_{\rm R} (1^{\circ}<|b|<3^{\circ})& h_{\rm z}(|b|<1^{\circ}) & h_{\rm z}(1^{\circ}<|b|<3^{\circ}) & 
   \overline{h_{\rm R}}(o)& \overline{h_{\rm z}} (o)& 
   \overline{h_{\rm R}}(y)&\overline{h_{\rm z}}(y) \cr
 & \degr & kpc& kpc& pc & pc  &kpc &pc&kpc&pc \cr
\noalign{\vskip2pt\hrule\vskip2pt}
{\rm A}& 0 & 0.25(0.04) & 0.25& 10(20) & 150     &0.2 &200 & 0.35& 30 \hfil\cr
{\rm B}& 6 & 0.95 & 0.75 & 90 & 350     & 0.75 & 250& 0.75 & 100 \hfil\cr
{\rm C}& 18 & \infty & \infty & (100) & (300) & 10.0&350  &\infty& (10) \hfil\cr
{\rm D}& 35 & 3.5 & 5.5 & 100 & 250 & 10.0 &200 & 1.5& 100 \hfil\cr
\noalign{\vskip2pt\hrule\vskip2pt}
}
\endtable

However, for all our measurements, except those at C (Fig.\SCAL),
this technique gave unique results (Table \HRZ, columns 3--6),
within the context of using double exponentials. 
In Table \HRZ\ the deprojected scales are also given
for the high-- and low--outflow sources separately,
determined over the whole latitude range of the survey.
For the galactic disk ($ |\ell| $\lsim 15\degr ),
most references (see Sackett 1997)
give $h_{\rm R}=$2.5--4.5 kpc and $h_{\rm z}=$250--400 pc (``thin'')
and $h_{\rm z}=$750--1500 pc (``thick'' disk).

\subsection {Ring ?}

The very large \arh\ at $\ell \sim -18^{\circ}$ (C; Fig.\SCAL)
indicates a flat distribution in longitude.
With this value of the measurement, the deprojection is not
unique. Nevertheless, our tests showed that 
projection effects alone, although being largest at these
longitudes, can not explain the extreme
\arh\ and we conclude it is intrinsic.
The flat distribution is formed mainly by the 
high--outflow (and single--peaked) 
sources, ie. the younger stars (Table \HRZ , column 7 vs. 9; 
Fig.\SCAL (circles)).
This region around $\ell = -18^{\circ}$ will be treated 
in more detail in \S 5, where we find that the distribution
in that direction is probably dominated by a ring structure.

\subsection {Central disk}

A deciparsec--scale, flat, rapidly--rotating flat
population of OH/IR stars 
is present in the galactic Centre (Lindqvist, Habing \& Winnberg 1992;
Sevenster \etal1995).
Even though the AOSP sample contains only 19 out of the 134 known OH/IR stars
in this region
(Lindqvist \etal1992; 52 more were discovered by Sjouwerman \etal1998a),
we can still resolve its very low \zh\ with respect to other
regions (Table \HRZ , column 5). 
As was known before,
the  central disk is seen primarily in the high--outflow
sources (Table \HRZ , column 10).

However, the \rh\ of 250 pc (Table \HRZ, column 3) is unlikely
to represent this small central disk.
Indeed, if we apply our method to the full sample of Lindqvist \etal(1992)
we find $h_{\rm z}=20$ pc and $h_{\rm R}=40$ pc, the same values
as they find themselves (Table \HRZ , columns 3,5 between brackets).

\beginfigure4
\fignam\VEXP
{\psfig{figure=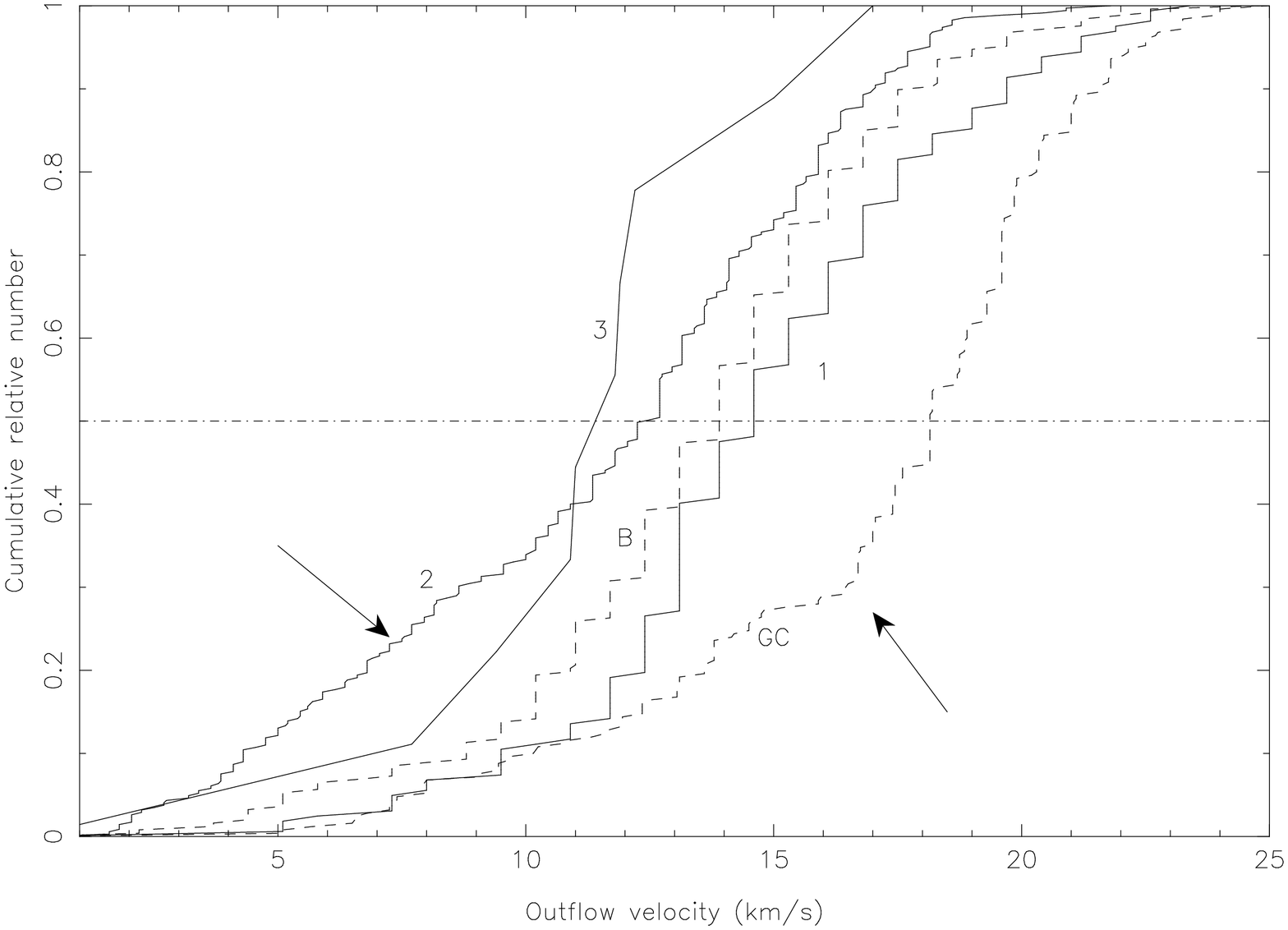,width=\hsize,angle=270}}
\caption{{\bf Figure \nfig}
The cumulative--number distributions of outflow velocities
for double--peaked OH/IR stars 
from various samples:
1. Disk AOSP sample;
2. Arecibo sample (Chengalur \etal1993);
3. Outer Galaxy (Blommaert \etal1993);
B. Bulge AOSP sample;
GC. galactic--centre sample (Sjouwerman \etal1998a).
The median outflow velocity correlates
with metallicity (equation (\MU)); its decrease with larger mean radius 
(in order : (GC,B) 1,2,3) reflects mainly the galactic metallicity gradient.
The small difference between ``1'' and ``B'' is mainly due to 
a mean--mass difference, however.
The arrows indicate deviant sub--populations of distribution
``2'' (thick disk; discussed in \S 3.4) and ``GC'' (star burst; \S 7). 
Except sample ``3'' that consists of 9 stars, 
all samples contain more than 200 stars.
}
\endfigure

\subsection {Outer Galaxy}

Hardly any OH/IR stars are known outside the solar circle.
Carbon--rich AGB stars, on the other hand, are hardly seen in
the inner Galaxy (see the figures in Blanco 1965). 
Blommaert, van der Veen \& Habing (1993) find that the few OH/IR stars
in the outer Galaxy have initial masses of at least 2--3 \msun ,
similar to the AOSP disk stars (\S 2). 
The low outflow velocities of the sample (the median is 
3 km/s lower than for the Disk AOSP sample, Fig.\VEXP ) 
indicate low metallicity (equation (\MU)), as was 
observed by Blommaert \etal(1993).
At intermediate longitudes ($\ell \sim 50^{\circ}$), 
the OH/IR stars from the Arecibo survey (Chengalur \etal1993)
have initial masses and ages similar to the Bulge AOSP sample
(\S 2) and slightly lower median outflow velocity.

The decrease in outflow velocities with higher longitude
is the result of decreasing metallicity.
The galactic--disk metallicity gradient ($-$0.07 dex kpc$^{-1}$ 
from oxygen, Smartt \& Rolleston 1997) corresponds, for constant mass,
to an outflow--velocity gradient of $\sim$1 \kms\ kpc$^{-1}$ 
(for \vexp$\sim$14 \kms, equation (\MU)).

At the same time, at lower metallicity, the limiting mass above which stars
remain oxygen--rich and below which carbon--rich stars form increases. 
The upper limit to the mass for which a star will reach the 
AGB in turn becomes lower with lower abundance (see Bertelli \etal1994).
The upper limit to the outflow velocities will decrease even
more noticeably (equation (\MU)); this explains the 
small \rh\ for the high--outflow sources (Table \HRZ).
The abundance may decrease rather suddenly around $R=6$ kpc,
as observed by Simpson \etal(1995).  They
explain this by assuming this is the outer edge of the zone--of--influence 
of the Bar. As we will see later (\S 6), the Bar's outer--Lindblad
resonance (OLR, Binney \& Tremaine 1987 (BT) Ch.6) may 
indeed have a radius of 6--7 kpc.

\beginfigure*5
\fignam\THICK
\hskip .5truecm{
\psfig{figure=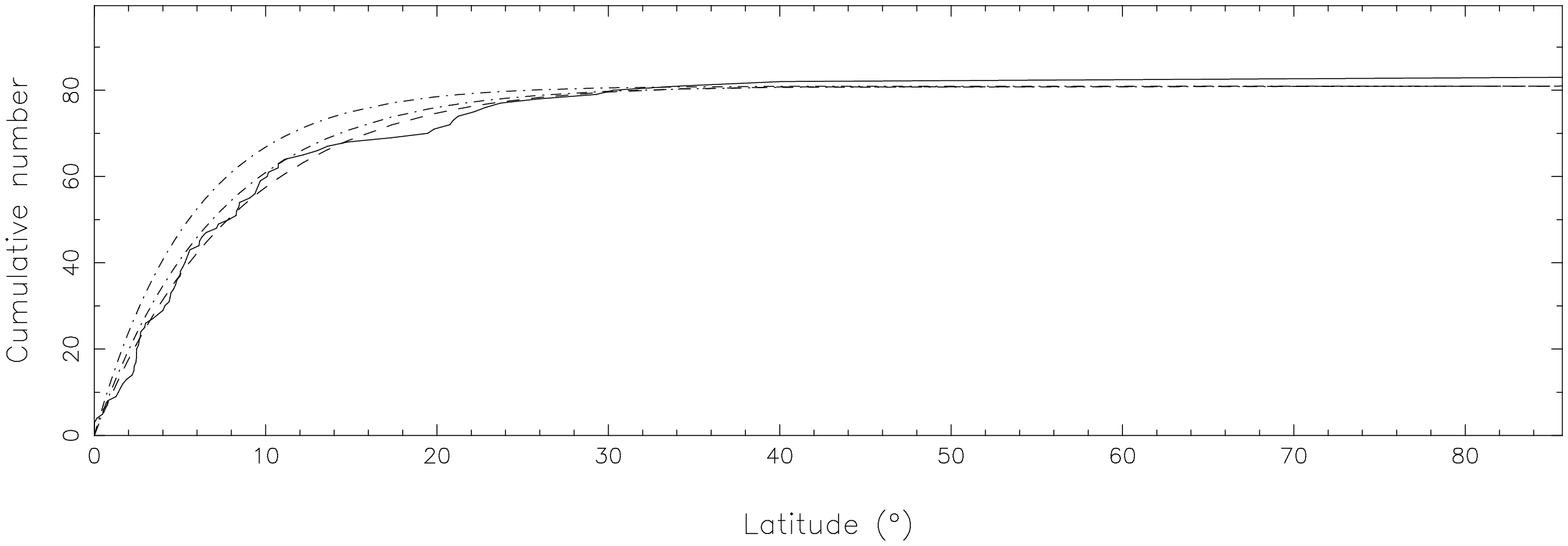,height=12truecm,angle=270}}
\vskip -6truecm
\caption{{\bf Figure \nfig} 
The cumulative--latitude distribution of all Arecibo stars
with outflow velocity $<$10 \kms\ (solid, cf.~Fig.\VEXP, \BDV ). 
The fit (dashed)
is a projected double--exponential distribution with 
\azh\ $h_{{\rm z}b}=$1.1 kpc. For comparison, the dot--dashed
lines give the distribution for $h_{{\rm z}b}$=800 pc and 1 kpc,
respectively. The intrinsic scaleheight is probably close to 1.1 kpc as well.
}
\endfigure

\beginfigure6
\fignam\BDV
{\psfig{figure=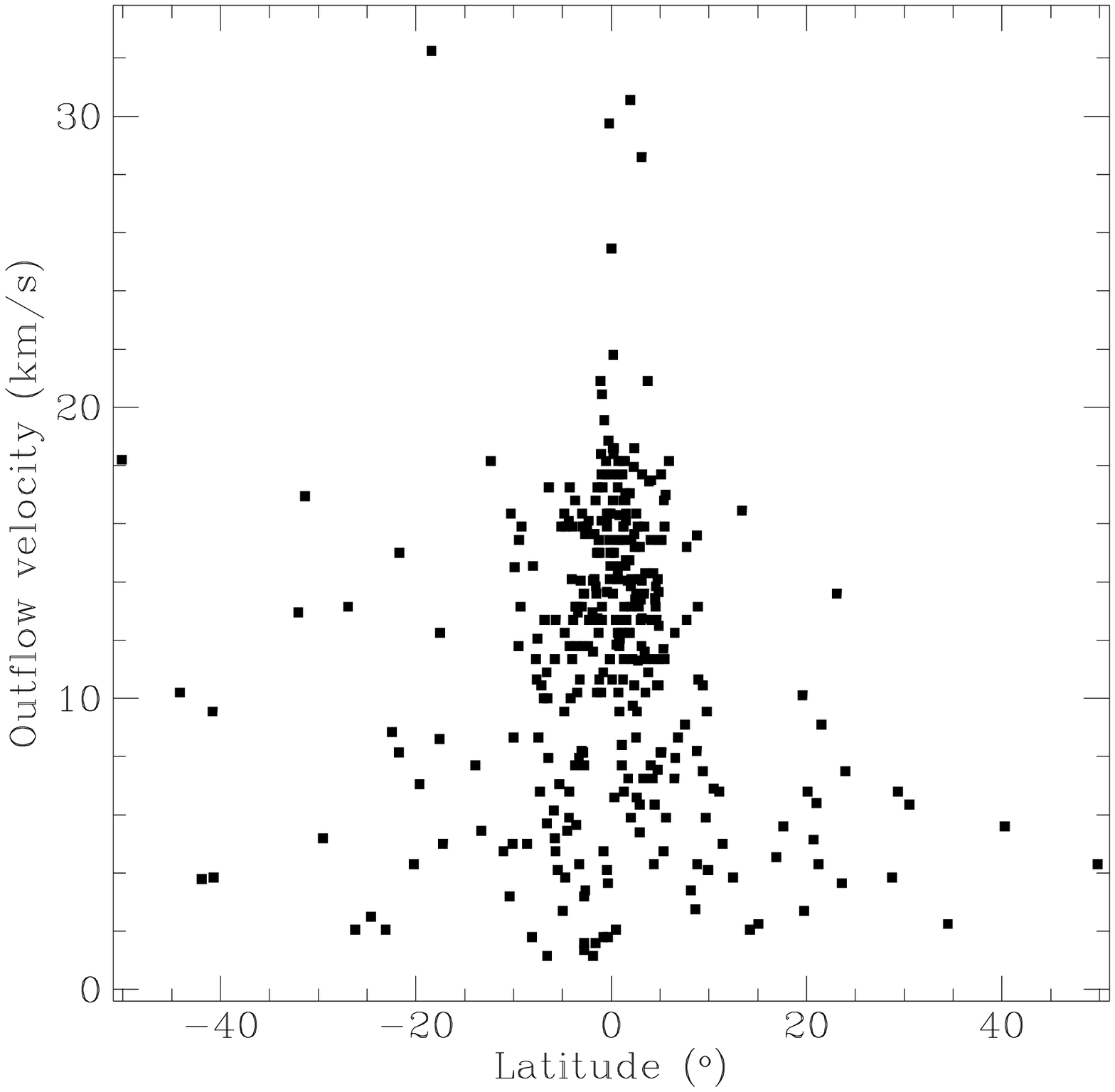,height=\hsize}}
\caption{{\bf Figure \nfig } 
The distribution of the Arecibo stars
over latitude and outflow velocity.
There is a difference in the concentration toward the plane
at low latitudes
between stars with outflow velocities higher and lower 
than 10 \kms, respectively. Below 10 \kms\ there is also
a clear excess of sources (50\%)in the Arecibo sample with
respect to the other samples in Fig.\VEXP. These stars
form possibly a thick disk.
}
\endfigure

\beginfigure7
\fignam\VEXPP
{\psfig{figure=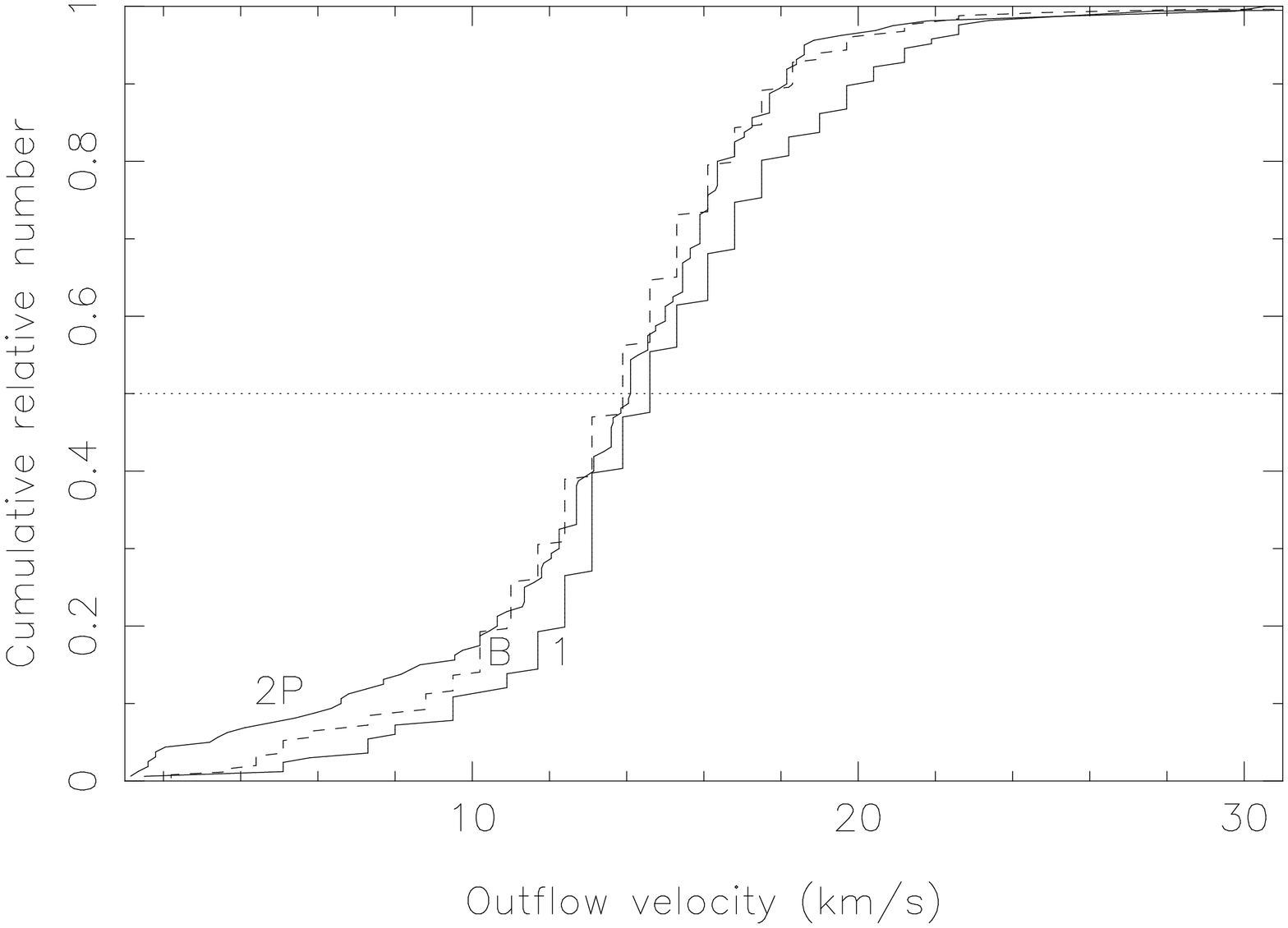,width=\hsize,angle=270}}
\caption{{\bf Figure \nfig } 
As Fig.\VEXP, for the Bulge AOSP sample (``B'', 250 stars in the plot), 
the Disk AOSP sample (``1'',165 stars)  and the
stars of the Arecibo sample with $|b| < 3^{\circ}$ (``2P'', 160 stars).
There is still an excess of sources below \vexp=10\kms\ for ``2P''.
Note that the distribution of higher outflow velocities of
``2P'' is much more like ``B'' than like ``1''.
}
\endfigure

\subsection {The Arecibo sample }

With respect to an otherwise equivalent sample
(te Lintel \etal1991), the Arecibo sample has twice as many
sources at outflow velocities \lsim 9 \kms\ (Fig.\VEXP, see 
left--most arrow ). This difference is significant 
to 99.8\% (Kolmogorov--Smirnov).
The apparent scales of the subset with 
outflow velocities $<$ 10 \kms\ (Fig.\BDV) are $h_{{\rm z}b}=$ 1.1 kpc
and $h_{{\rm R}\ell}\rightarrow\infty$ (Fig.\THICK). 
Again, the nearly--infinite \arh\
makes unique deprojection impossible, but we found that
the \azh\ is most probably close to the true \zh\ in this case.
These stars might be part of the 1--kpc
thick disk claimed by other authors (Habing 1988; 
Gilmore \& Reid 1983, Ojha \etal1996), although
Blommaert \etal(1993) conclude that true AGB stars
do not partake in the thick disk and there is no sign of this
population in the te Lintel (\etal1991) sample.
In Fig.\BDV, we see that these low--outflow
stars are not concentrated toward the plane. This may be 
a selection effect as the Arecibo sample was acquired in a detection
experiment toward IRAS--selected point sources, but this is also
true for the te Lintel (\etal1991) sample. Moreover,
when we select stars from the 
Arecibo sample in the same latitude range as the AOSP sample
a small excess is still present (Fig.\VEXPP). 

So, in the Arecibo sample we see an extra population with
low outflow velocities and a gap in the latitude distribution.
The latter brings to mind the intriguing ``levitation''
process (Sridhar \& Touma 1996a,b), a vertical--to--radial
resonance. This would not explain an excess of sources, however. 
If the excess sources have low metallicity,
rather than only low masses (equation (\MU)), this population 
could have been accreted anytime during the last 10 Gyr from
a dwarf such as Fornax ([Fe/H] $<-0.7$).
With the observational limit of less than five 
such more metal--rich dwarfs accreted in the last 10 Gyr 
(Unavane, Wyse \& Gilmore 1996) this is a possibility.

\subsection{Radial and vertical variations}

Clearly, one double--exponential cannot describe the distribution
of OH/IR stars. The older disk stars (few Gyr)
seem to trace an exponential disk with rather large scalelength
plus a central component with similar vertical scale (Table \HRZ).
The younger stars ($<$ 1Gyr) follow a more capricious pattern
and have varying, but small vertical scales.
This is not unexpected since they have completed only a few 
rotations around the \gc\ since their formation and therefore
phase mixing, let alone radial mixing, has not yet been effective.
There seem to be no sudden transitions in the distribution;
except that from the flat distribution ($h_{\rm R}\rightarrow \infty$)
to the outer disk ($h_{\rm R}=$1.5 kpc)
for the high--outflow sources 
that is rather abrupt. As we will see later (\S 5), this may be
around the radius of corotation with the pattern speed
of the central Bar.
The radial scale derived (via the same procedure described earlier)
from the Arecibo sample (using only $|b| > 8^{\circ}$ because of
it's incompleteness in the plane) and from 
a sample of carbon--rich AGB stars 
($|\ell| \sim 60^{\circ}$, $|b| < 15^{\circ}$, Loup \etal1993),
is $\sim 10$ kpc; the same as from the low--outflow OH/IR stars.
So, except for the massive oxygen--rich objects,
the scalelength is found to be similar for all AGB stars.
The very different radial distributions of C--rich and 
O--rich AGB stars, respectively, is governed by the metallicity
gradient only.

Outside the AOSP--survey window, we also derived scaleheights
from the Arecibo subsample and the Loup sample, and found
$h_{\rm z} \sim 500$ pc for both.
The full sample of outer--Galaxy AGB (not only
OH/IR) stars of Blommaert \etal(1993) has $h_{\rm z}$ \lsim 400 pc. 
The vertical scale appears to increase with latitude
from $\sim$100 pc to $\sim$500 pc, irrespective of the longitude range.
The increase in scaleheight within the Disk AOSP sample agrees
with the diffusion models by Wielen (1977) for age increasing
from \lsim 1 Gyr to \gsim 2 Gyr. The scaleheight of 500 pc for 
the oldest AGB stars is the same as that for white dwarfs as
given in Mihalas \& Binney (1981).

\subsection {In short}

In summary, the AGB stars are distributed in the thin (old) disk with 
a scaleheight of 100 pc for the youngest AGB stars (\lsim 1 Gyr).
The scaleheight increases continuously 
to 500 pc for AGB of \gsim 5 Gyr. 
The scaleheights for the Disk and the Bulge are the same.
The carbon--rich and oxygen--rich AGB stars form
one population; the differences between the distributions of the
two groups are governed mainly by the metallicity gradient.
There seems to be a small population that has a distinctly large
scaleheight.

All trends in the 
AOSP sample can be seen in Fig.\SURF\ in a pictorial fashion.
The exact numbers for the scales in Table \HRZ\ should be used 
with care, since our deprojection method is indirect.
Nevertheless, from modelling distribution functions for various
of the samples used here the same trends
numbers emerge (\CHFOU). Also, the scales in the Bulge region
agree very well with those recently found from DENIS data 
(Gilmore priv.comm.).
%2 kpc for high, 5 kpc for low, 5 kpc for carbon Schechter

\section{The bar}

\subsection{Surface density}

In this section, we compare parametrized bar models to
the surface density of the AOSP sample. 
We do not optimize fits quantitatively, since the choice of a 
parametrized bar model is already an arbitrary one. 
Rather, we take bar models that are known to give
good approximations to other observations and see how well they
agree with ours.

\beginfigure*8
\fignam\BARG
\vskip .5truecm
\hskip .05truecm{\psfig{figure=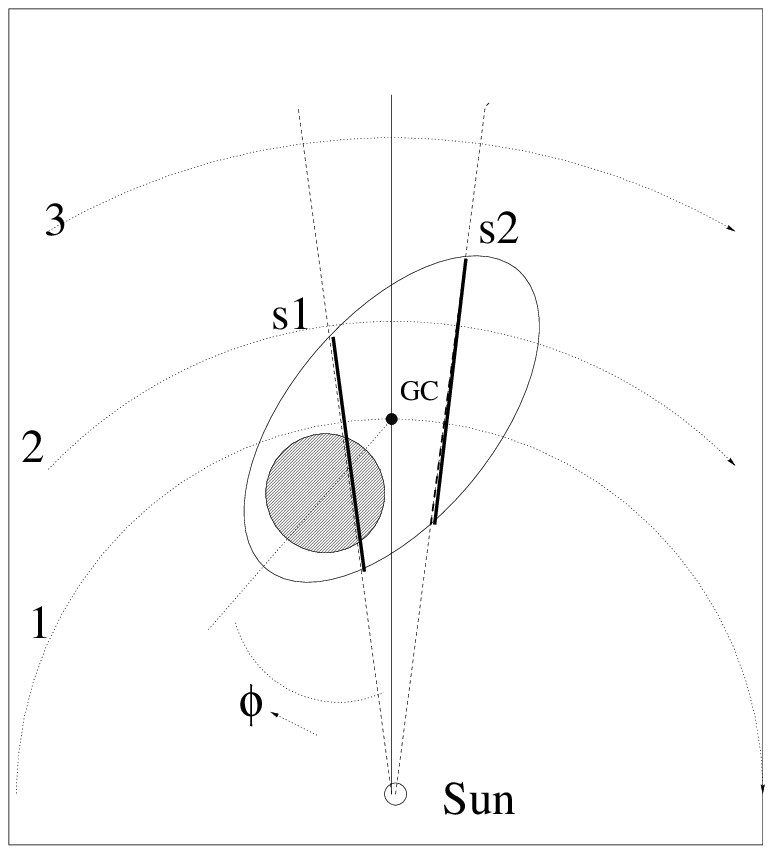,height=4.6truecm}}
\vskip -5.25truecm
\hskip 3.65truecm{ \psfig{figure=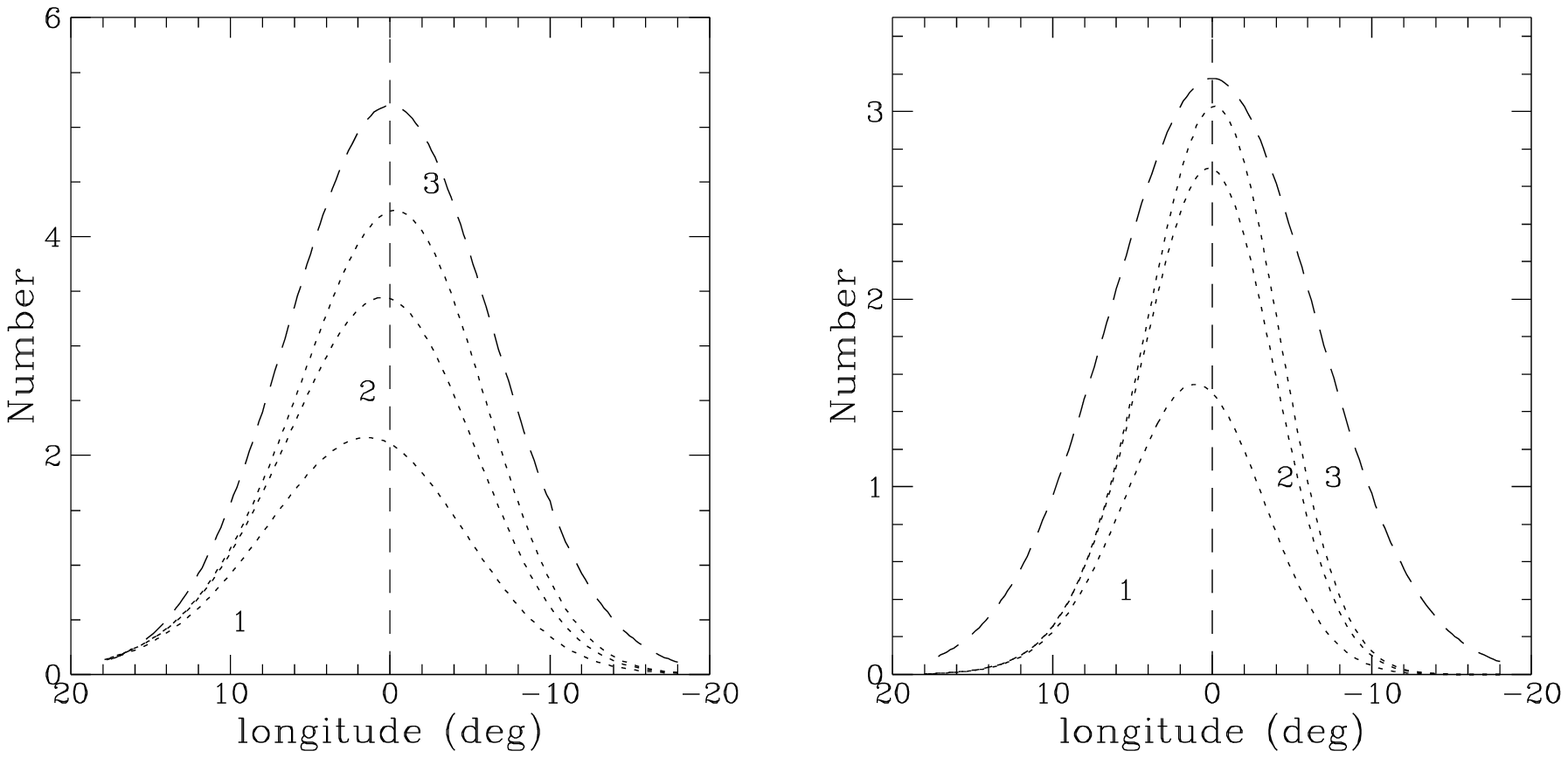,height=11.7truecm} }
%%ngaus
\vskip -6truecm
\caption{{\bf Figure \nfig} 
{\bf a} (left) 
Schematic view of the Bar from the galactic North pole. The viewing
angle $\phi$ is indicated; the near end of the Bar is most certainly
at positive longitudes. The lines of sight (s1 and s2) at equal
but opposite longitudes differ in length.
The dotted arcs show schematically possible observation limits.
The shaded
circle represents a lopsided distortion, that would project
to a similar surface--density distribution as the Bar for 
an observation limit of \rsun\ (dotted arc 1).
In {\bf b} and {\bf c} we show 
the total number of stars (arbitrarily normalized) along the line of sight 
in a two-dimensional elliptical barmodel with gaussian density distribution
for three different integration limits (8 kpc (1), 9 kpc (2) and $\infty$ (3); 
$R_{\odot}\equiv$8 kpc).
The integration limits correspond to the observational limits 
shown in {\bf a}. 
For reference a gaussian is shown (dashed). Note the offset of the maximum
toward negative longitudes for the largest integration limit.
{\bf b} (middle) Axis ratio $q$=0.6, semi--major axis $a$=3.5 kpc 
(=3$\sigma_{\rm gaus}$).
{\bf c} (right) Axis ratio $q$=0.4, semi--major axis $a$=2.5 kpc 
(=3$\sigma_{\rm gaus}$). 
The viewing angle $\phi$ is 20\deg\ in both panels {\bf b} and {\bf c}.
The model in {\bf c} is based on the K--band G2 model 
for $R_{\rm max}=2.4$ kpc of Dwek \etal(1995).
}
\endfigure

The viewing angle $\phi$ is the angle between the 
Bar's major axis and the \losn\ toward the \gc\ (Fig.\BARG (a)).
In Fig.\BARG (b,c) we show the surface density distribution as a function of
longitude, $ N(\ell)$, for two 
flat (two--dimensional) elliptical bars with gaussian density distribution
(cf. G2 model in Dwek \etal1995).
The viewing angles are taken to be 20\deg, as suggested by
some observations (Dwek \etal1995). 
The form of $ N(\ell)$ depends upon the distance out to which
we integrate or, in observational terms,
the distance $d$ out to which the data sample the Galaxy, so
 $  N(\ell) =  N(\ell; d)$.
For values of $d \sim $\rsun, $ N(\ell; d)$ essentially
looks like the distribution arising from an $m=1$ distortion 
(lopsided; see Fig.\BARG (a)),
with its maximum toward {\it positive} longitudes.
With $d = \infty$, however, the distribution
is skewed toward {\it negative} longitudes.
This is the result of the line of sight through the $m=2$ distortion
being longer on the far side than on the near side, for
small values of absolute longitudes (Fig.\BARG (a)). This effect was first
predicted by Blitz \& Spergel (1991) and is also seen
in the micro--lensing maps calculated by Evans (1994).
The strength of this effect obviously depends upon the parameters
of the density distribution of the bar.

We will try to find this effect in our Bulge AOSP sample.
The \dpk\ OH/IR stars are divided into two equally--sized samples
by outflow velocity, at \vexp=14 \kms, with
average \vexp\ of 11.3 (sample I) respectively 18.3 (sample II) \kms\ 
(cf. Fig.\SURF (e,d)).
This gives a factor of 1.7 difference in stellar luminosity $
L_{\ast}$, even if we assume $\rm \mu_{I}=2 \mu_{II}$ (equation (\MU)).
Blommaert \etal(1997) find a range in $\mu$  of $\sim$ 2 
in the GC with
IR observations. The range of masses in the Bulge (\S 2)
makes a factor of 1.7 difference in $L_{\ast}$ ($\sim$1.5 in $M_{\rm i}$)
very well possible.
Since the OH masers are saturated, the OH luminosity $ L_{\rm OH}$
increases, on average, linearly with $ L_{\ast}$.
The lower limit to the flux density $ S_{\rm OH}$
is the same for both samples,
so the average limiting distance of sample II
is a factor $\sim 1.3$ larger than of sample I.
We can thus use the two samples to mimic the different
integration limits in Fig.\BARG\ (see also \S 2).

\beginfigure*{9}
\fignam\BARF
\hskip 0.5truecm{
\psfig{figure=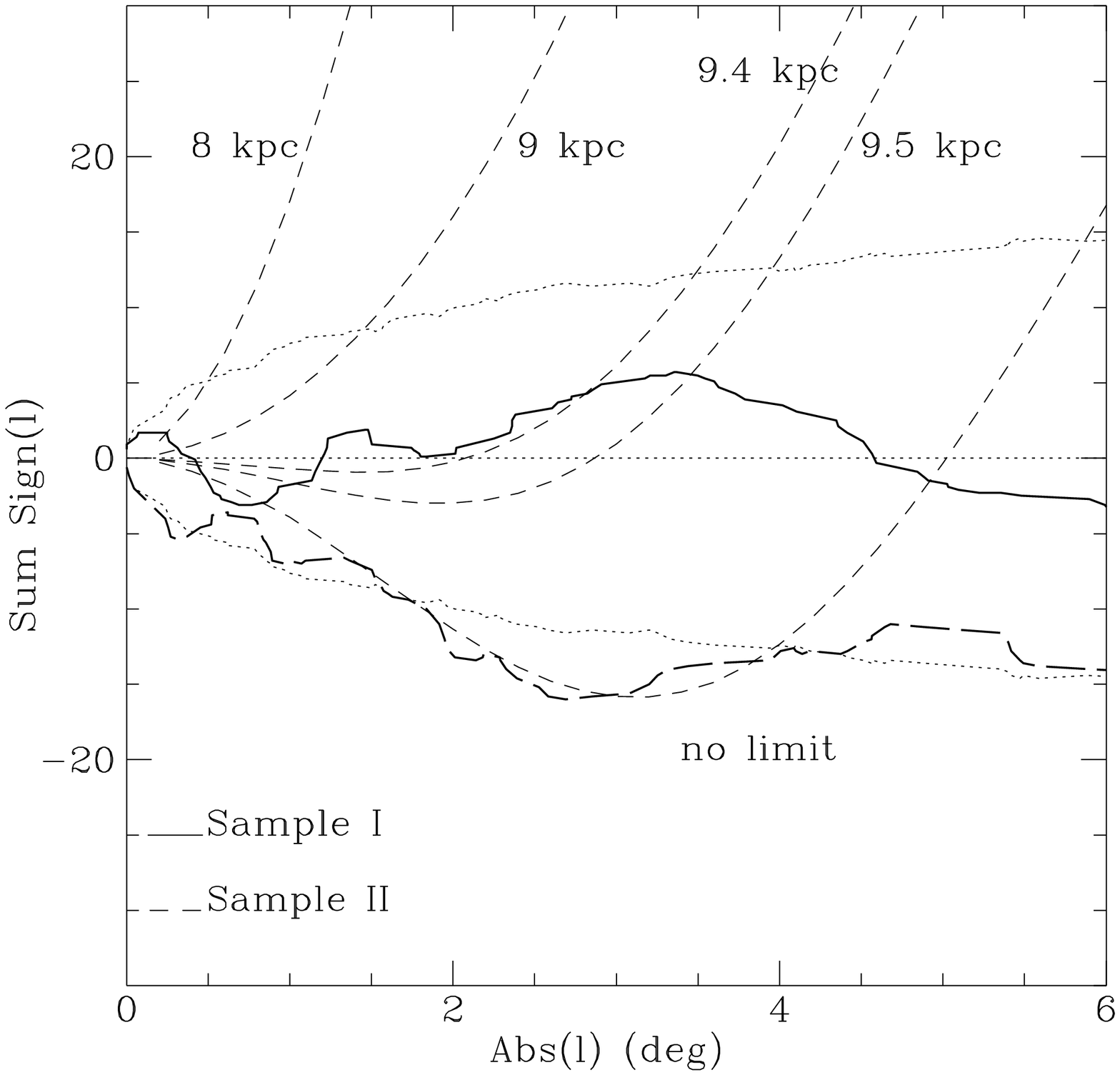,height=7.5truecm}
}
\vskip -7.5truecm
\hskip 8.0truecm{
\psfig{figure=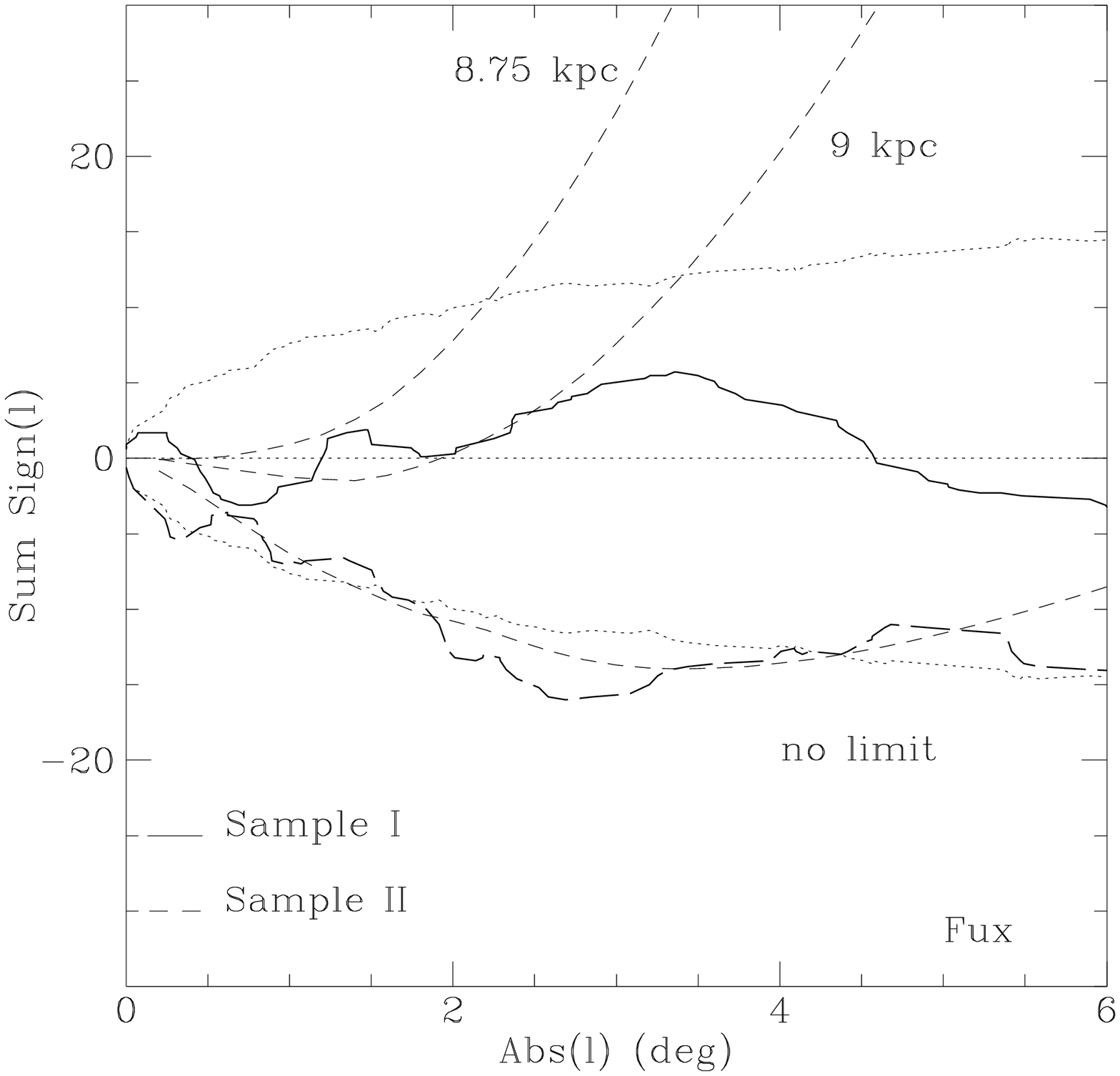,height=7.5truecm}
}
%%nmod
\caption{{\bf Figure \nfig} 
{\bf a} (left)
The cumulative sum $\sum (\ell / |\ell|)$ versus $ |\ell| $ after
sorting on $|\ell |$ for the two samples (thick solid and dashed lines) and
for the G2--bar model from Fig.\BARG (c)
with different integration cut-offs (short dashed lines, see Fig.\BARG ).
The dotted curves indicate the values for which
the probability of the sum arising in a binomial distribution being
larger (or smaller for negative values) than that value is 5\% 
(ie. single--sided).
For $|\ell| < 0.5^{\circ}$ and
$ |\ell| > 5^{\circ} $ contributions by additional 
features influence the distribution (see Fig.\SURF ; \S 5).
{\bf b} (right)
Same as {\bf a}, but for a bar model consisting of two exponential
elliptical--bar profiles with major--axis scalelengths of 
200 pc and 1 kpc (central--density contrast 8.5),
with viewing angle of 45\degr\ and axis ratio of 0.5. 
The profile is an approximate fit
to the density in the plane of 
an N--body bar and the viewing angle of 45\degr\ optimizes the fit between
the AOSP data and the N--body model (\CHTHR ; Fux 1997).
}
\endfigure

The effect of skewed distributions should be clearest in the
inner regions of the Galaxy (see Fig.\BARG (b,c)).
The ratios of the number of stars with $ 0^{\circ} < \ell < 4^{\circ} $
to the number of stars with $ 0^{\circ} > \ell > -4^{\circ} $
are 39/35 (sample I) and 22/35 (sample II).
These ratios are in accordance with the theoretical results shown
in Fig.\BARG .
To define these trends in a more sophisticated way, we sorted both samples
on {\it absolute} longitude and 
calculated the cumulative sums of the sign of the longitude
$ \sum ( \ell / |\ell| ) $;
we add or subtract 1 for each
star (Fig.\BARF). An axisymmetric distribution gives a line that hovers
around zero. If  negative (positive) longitudes are
\lq overpopulated\rq\ the sum will steadily decay (rise).
This sum is shown in Fig.\BARF\ for the two data sets and for
the bar model shown in Fig.\BARG (c), as well as for 
an N--body model (Fux 1997) found to represent the AOSP sample well
(\CHTHR ).  The dotted curves give
the 95\% confidence limits for deviation from axisymmetry.
Sample I (solid curve) never deviates significantly from
axisymmetry, although there are local trends similar to those of the
models with the intermediate cut-off.
Sample II (dashed curve) , however, lies at or outside the 5\% confidence
level and coincides remarkably well with the
models without distance cut-off.
The set of stars with
low \vexp\ would have an average distance cut-off of 
9--9.4 kpc according to these models, 
the set with high \vexp\ of around 12 kpc, beyond
which there is no significant contribution from the bar to the
integrated density anymore. This agrees very well
with the difference of a factor 1.3 in
average distance derived from the relation
between \vexp\  and stellar luminosity.

We estimate the disk contamination to be \lsim 20\% within
5\degr\ of longitude from the GC (\S 2; \CHFOU). 
If an axisymmetric
component contributes significantly in the inner degrees,
the evidence for the existence of the Bar would only become stronger.
The $\sum (\ell/|\ell|)$ distribution would not change
with the subtraction of a projected axisymmetric distribution
of any relative density, but $N(\ell < \ell_{lim})$ would become 
lower. This would make the probability of the deviation coming from
a binomial distribution even smaller. In other words, the dotted curves
in Fig.\BARF\ would shift horizontally 
to the right, but the data--lines would remain
in place on average.

The evidence presented here is the first 
large--scale morphological evidence for a
\gba\ that cannot also be explained by a physical lopsided
density distribution (Sevenster 1996).
The effect is also clearly visible in Fig.\SURF (b,c) (sample II,I)
in a qualitative way. The same trend in the
longitude distribution is reported by Unavane \& Gilmore (1998)
who use L--band data in the plane. However, the 
necessary extinction corrections are estimates and of similar order
as the asymmetries one is after.
Unavane \& Gilmore (1998) find that the Dwek models (E2, G3)
give too large a deviation from axisymmetry, whereas 
the G2 model gives a good representation of our data (Fig.\BARF (a)).
The results in Fig.\BARF\ demonstrate that only for a given
density model the parameters of a bar can be optimized
(Zhao 1997), as both rather different models give 
good representations of the data. 

\subsection{The Bulge's kinematic type }

\beginfigure*{10}
\fignam\VELO
\vskip -5truecm
\hskip 0.5truecm{\psfig{figure=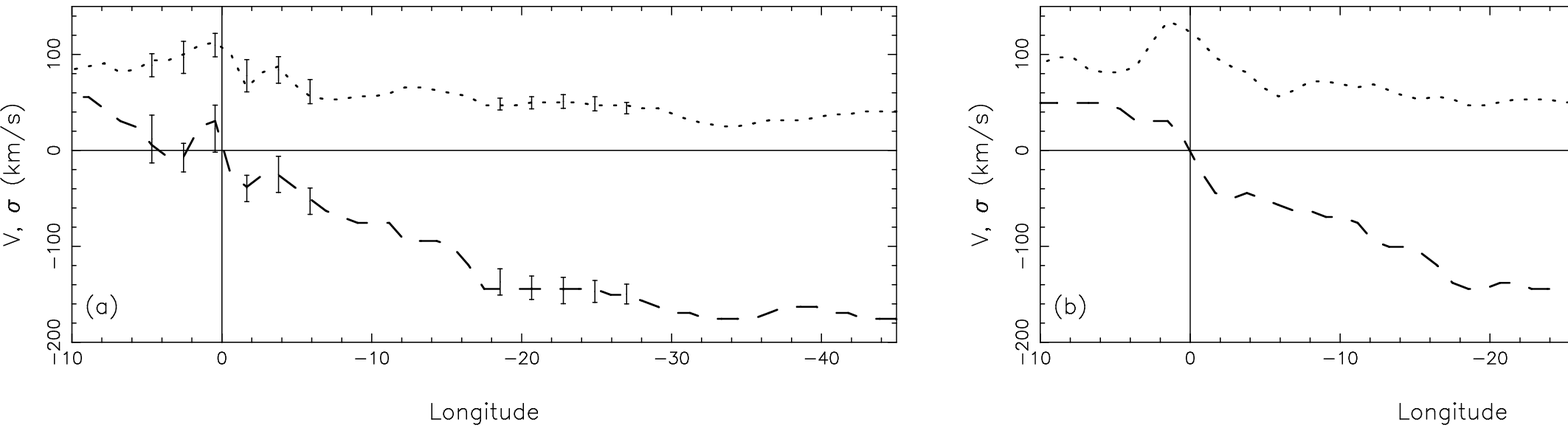,height=\hsize}}
\vskip -8truecm
%%rotsig
\caption{{\bf Figure \nfig} 
The mean \losa\ velocity (dash) in the inertial frame (for $V_{\rm LSR} 
\equiv $200 \kms)
smoothed data (double--peaked stars only) at $ b = 0^{\circ}$ 
({\bf a}) and 
$ b = 2^{\circ}$ ({\bf b}). The gridsize is 1\degr\ by 6 \kms,
the initial kernel sizes are 1\degr\ and 30\kms, respectively.
We show errors derived via bootstrapping (Press \etal1992, see \CHFOU)
for the features that we discuss in the text.
(The mean velocity is the 50\% value and dispersion is half the
difference of the 83\% and the 17\% values
of the smoothed velocity profiles.)
}
\endfigure

\noindent
The anisotropy parameter (Binney 1976) is a measure of the rotational support
of a system. It is the ratio of
the maximum rotational velocity of a bulge ($V_{\rm m}$) to its
central velocity
dispersion ($\sigma_0$). In combination with the 
flattening $\epsilon$, it provides a way to distinguish between 
rotationally--supported, 
dispersion--supported and streaming--dominated
systems (Illingworth 1977). The maximum
rotation velocity of the Bulge occurs at $\ell \sim -18^{\circ}$, where
the mean--velocity curve levels out (Fig.\VELO), so  
$V_{\rm m}$ = 140$\pm20$\kms .
For $\sigma_0$, we use 153 \kms\ (Blum 1996), which is
more reliable than our determination of this second--order moment
(135$\pm20$ \kms , Fig.\VELO(b)).
This gives a value of $V_{\rm m} / \sigma_0$
of 0.9$\pm$0.1 . Together with an ellipticity 
of $\epsilon \sim 0.4$
(Table \HRZ ; Dwek \etal1995, G0-model; Kent 1992), 
this locates the \gba\
in the $V_{\rm m} / \sigma_0$ -- $\epsilon$
diagram between
the oblate and the SA{\underbar B} bulges (Kormendy 1993),
governed by somewhat--more--than--rotational support.
The Galaxy fits well in the relation as an average SAB galaxy.

Earlier values for the Bulge's dispersion, 
from measurements toward Baade's window,
were much lower ($113^{+6}_{-5}$ \kms; Sharples, Walker \&
Cropper 1990). Adopting this as
$\sigma_0$, the Bulge would be located in the region
of extremely triaxial bulges 
in the $V_{\rm m} / \sigma_0$ -- $\epsilon$ diagram. 
Baade's window ($b=-$4\degr) is too far from the plane to 
sample the central Bar dispersion. 
We conclude that the Bar is so flat that
it cannot be assumed to be the dominant contributor 
to the distribution along the \losn\ toward Baade's window (Table \HRZ; 
see also \CHTHR). Paradoxically,
using the dispersion in Baade's window one would obtain
an anisotropy parameter indicative of a strong bar.

\section{Resonant structures}

In this section, we will concentrate on 
local features in the distributions of various tracers. 
For the AOSP sample, these structures are seen in Fig.\SURF ,
marked as \RRmi, \RRmu\ and \RRmc .
We complement our own evolved--stellar data set with 
2.4--GHz--continuum observations (Duncan \etal1995) and a sample of 
star--forming regions (Comeron \& Torra 1996).
The synchrotron radiation dominating the continuum at 2.4 GHz,
is found to be a good tracer of the current locations of density 
waves (eg. Tilanus \& Allen 1989). 

\beginfigure*{11}
\fignam\RADIO
\vskip -5truecm
\hskip -6pt{\psfig{figure=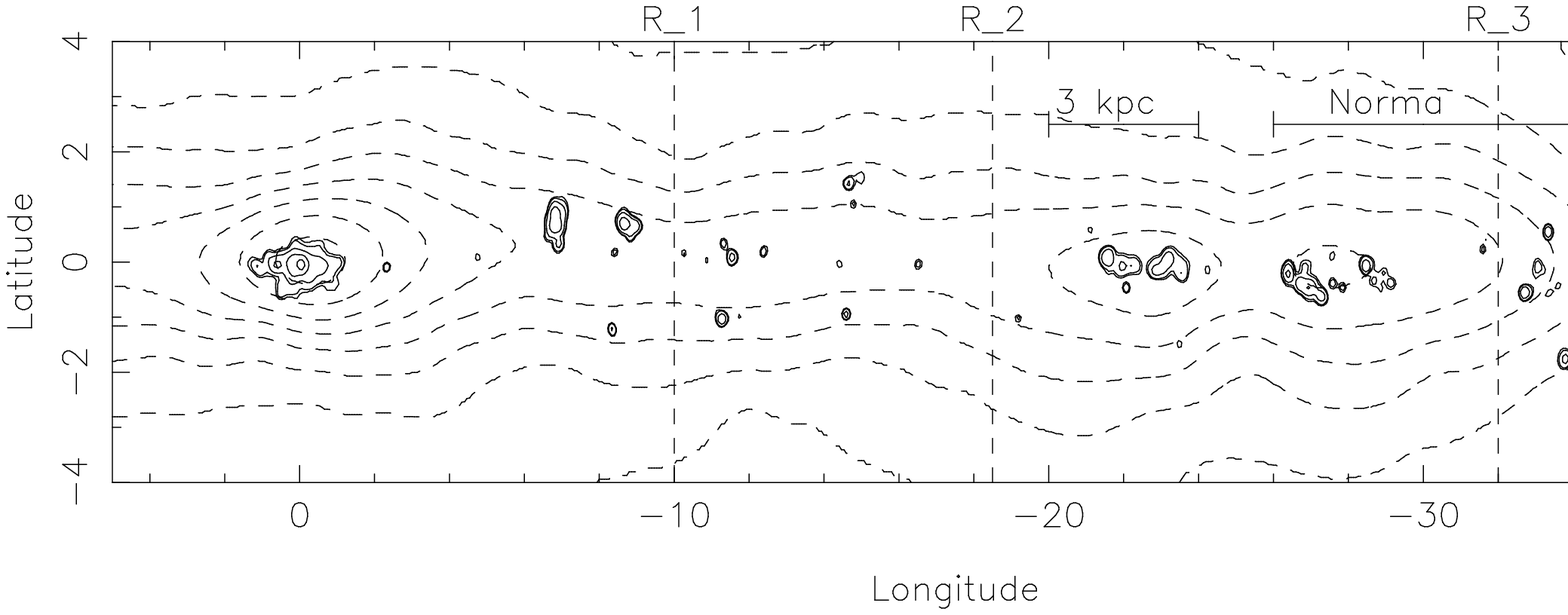,height=\hsize}}
\vskip -6truecm
\caption{{\bf Figure \nfig }
The observed surface--density distribution at 2.4 GHz due to
Duncan \etal(1995). 
The features \RRmi, \RRmu\ and \RRmc\ are indicated (see Fig.\SURF )
as well as the ``Norma arm'' (tangent point $-$32\degr) and the 
``3--kpc arm'' (tangent point $-$22\degr).
For the large--scale ($>$ 1\degr) distribution, the ten dashed contours are
spaced linearly between 10\% and 100\% of the maximum (large--scale) density.
For the small--scale ($<$ 1\degr) distribution, the ten solid contours are
spaced logarithmically between 3\% and 95\% of the maximum
(small--scale) density. The large--scale emission is a good tracer of
density waves.
}
\endfigure

Duncan \etal(1995) present their data separated in a small--scale--
(sub--degree) and large--scale distribution (Fig.\RADIO ).
The small--scale emission arises largely in 
supernovae; the large--scale emission traces
the molecular gas and density waves.
For the uninitiated, an introduction to non--axisymmetric
galactic dynamics can be found in Binney \& Tremaine (Ch.6), 
and an introduction to terminolgy and literature in Sevenster (1997).

\subsection{The corotation region}

The only direct method to find corotation (CR), via the pattern speed, 
is in most cases inapplicable (see Tremaine \& Weinberg 1984;
Merrifield \& Kuijken 1995). However, 
a number of indirect indicators can be used.
A signature of CR is often observed
in the form of dust lanes, spiral--arm
bifurcations or local decreases in the density (eg. Elmegreen 1996).
A flat part of the rotation curve may outline roughly the region
between the inner--ultra--harmonic resonance (IUHR) and 
CR (Wozniak \& Pfenniger 1997).
This is observed in the edge--on S0--galaxy
NGC$\,$4570 (van den Bosch \& Emsellem 1998).
Similarly, the orbital stochasticity in this 
region (Contopoulos \& Grosb\o l 1986) will cause efficient radial mixing,
smearing out the density gradient (see also Fux 1997).
An easily observed result of the dynamics
in this region are ``inner rings'' that are prominent in 
the majority of barred galaxies (see Buta 1996).

All these features are in fact present in the Galaxy.
The narrow gap between the so--called 3--kpc and Norma arms
(Fig.\RADIO ), also seen in other tracers (eg.~CO, Bronfman 1992),
could well indicate CR.
In good agreement with this is that the flat part of the rotation
curve found for the AOSP sample (Fig.\VELO ) stretches from 
$\ell \sim -18^{\circ}$ to $\ell \sim -25^{\circ}$. Those 
longitudes would, according to Wozniak \& Pfenniger (1997),
indicate IUHR and CR, respectively. The very flat
density found for this region in \S 3 further reinforces
this picture (Contopoulos \& Grosb\o l 1986). It also rules
out the possibility that the rotation curve is flat due to
a logarithmic potential (density $\propto R^{-2}$).

\beginfigure*{12}
\fignam\HIS
\vskip -2truecm
\hskip -2.5truecm{
\psfig{figure=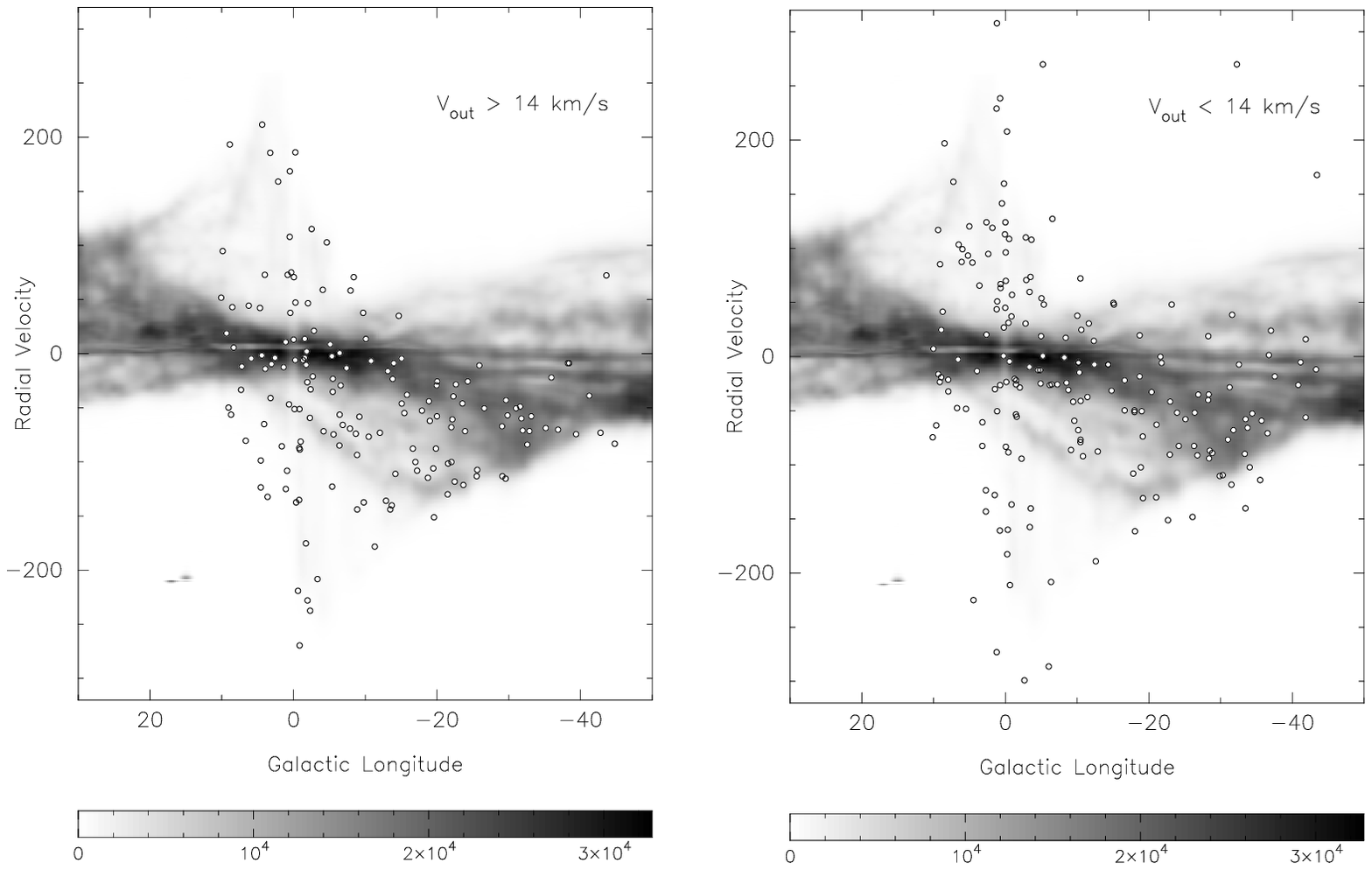,height=1.7\hsize}}
\vskip -0.9\hsize
\caption{{\bf Figure \nfig}
The \lvd\ for HI (grey scale) in the plane
and AOSP stars with $|b| <$ 1.5\degr.
Overplotted in the left panel are stars with 
\vexp $>$ 14 \kms\ (the young sample) and in
the right panel stars with \vexp $<$ 14 \kms\ (the old sample
plus the single--peaked stars).
The 3--kpc arm is clearly seen in the HI as the dark linear feature 
extending from ($+$5\degr, $-$20 \kms) to ($-$20\degr, -130 \kms).
The young stars have a subgroup of nine massive stars that follows
exactly the longitude--velocity structure of
this arm. The old sample shows no connection at all,
and deviates in general much more from the HI \lvd.
(Courtesy A. Kalnajs)
}
\endfigure

Most importantly, we argue that the 3--kpc arm is the projection not
of a spiral arm but of an inner ring, such as mentioned above,
for the following reason. A subsample of the AOSP sample follows
exactly the longitude--velocity structure of the
3--kpc filament (Fig.\HIS ). 
A group of nine young (high--outflow) stars, that trace the
kinematic structure between $0^{\circ} > \ell  > -10^{\circ}$,
stands out in the left panel of Fig.\HIS .
These nine stars are at very low latitudes and have
a very high median outflow velocity of 17.5 \kms. If the metallicity
is the same as in the rest of the disk, the initial masses 
of this 3--kpc sample would be $\sim$ 6 \msun\ and the
ages $\sim$ 600 Myr (Appendix A).
As those stars remain close to the gas for several galactic years,
their trajectories must follow closely that of the gas,
In other words, the gas filament must outline closed orbits
instead of a (temporary) spiral density--wave maximum.
With proper motions for these stars one could constrain the 
motion along the 3--kpc arm completely, something that is
impossible to achieve with gas.

\beginfigure*{13}
\fignam\AILR
\hskip 0.25truecm{\psfig{figure=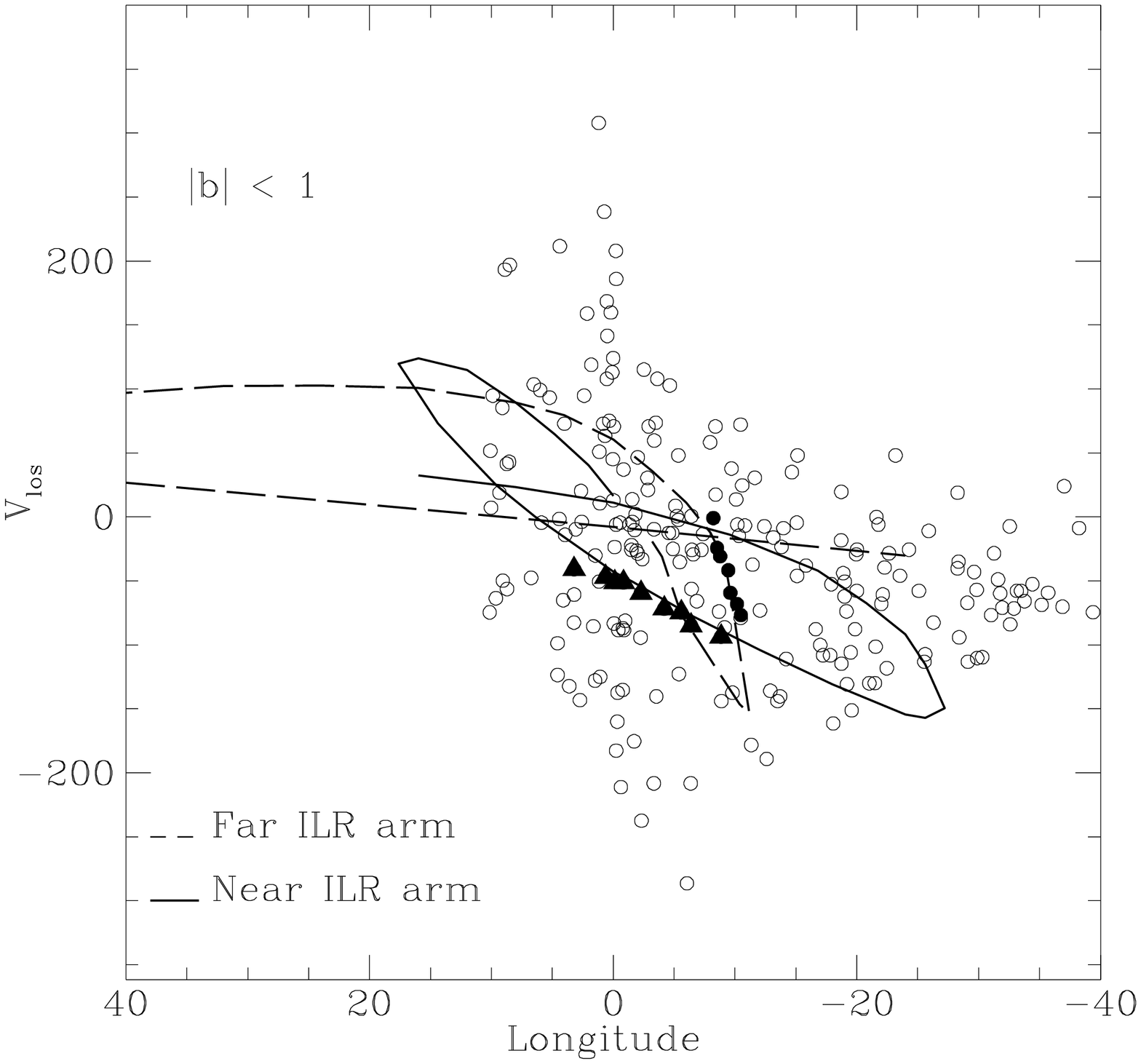,width=8.5cm}}
\vskip -8.1truecm
\hskip 8.5truecm{
{\psfig{figure=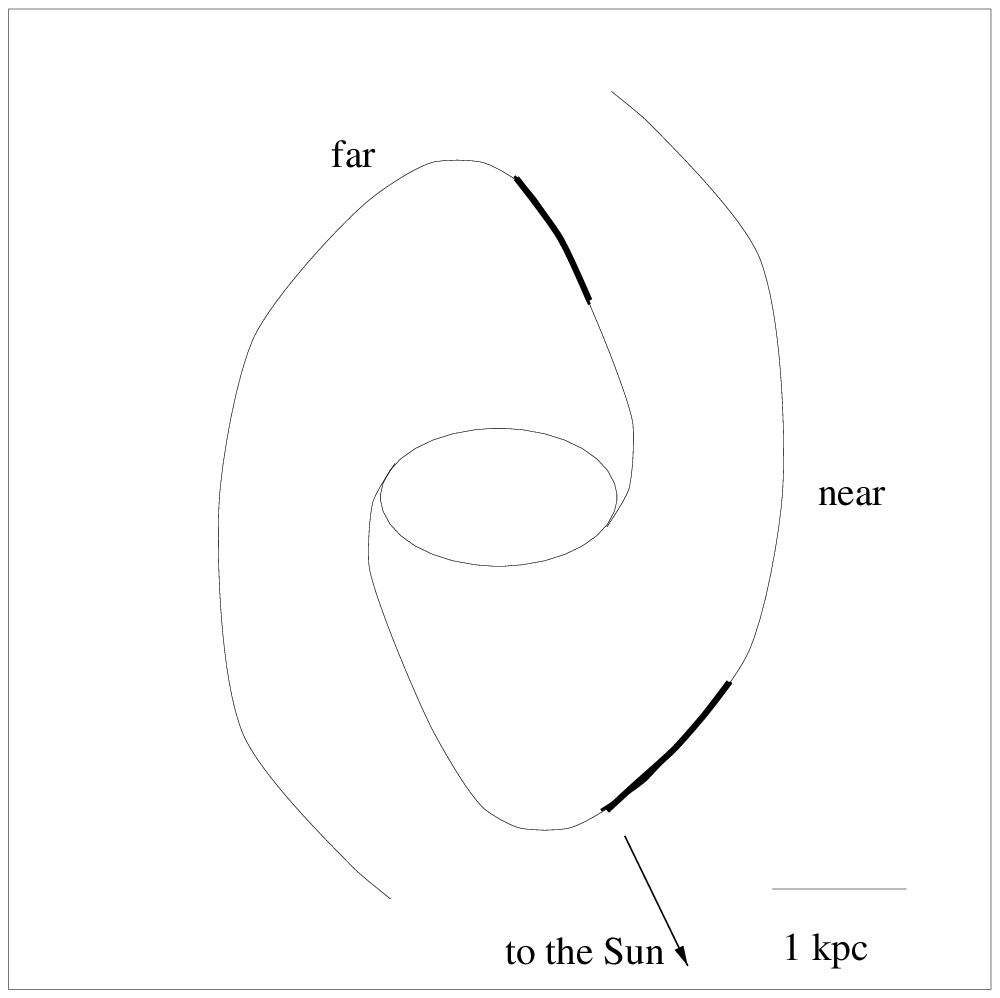,width=7.4cm}}
}
\vskip .6truecm
\caption{{\bf Figure \nfig}
{\bf a} The \lvd\ for AOSP stars with $|b| < $1\degr\ (open circles). 
The filled circles are the 7 stars from \RRmu\ with
$|b| < $ 1\degr\ and the 
filled triangles the 9 stars connected to the 3--kpc arm.
The solid curve is the ``near ILR arm''
from a model by Mulder \& Liem (1986) with $\phi$=20\degr\ and 
$R_{\rm CR}$=0.98\rsun ,
the dashed curve the ``far ILR arm'' (see {\bf b}).
The velocities are in the inertial frame
(changed from Mulder \& Liem (1986) for
\vlsr=200 \kms\ instead of 250 \kms.
{\bf b} The ILR arms from Mulder \& Liem viewed from the galactic North pole.
The thickened parts indicate roughly the positions of the two
groups of stars marked in {\bf a} if they are connected to those 
spiral arms. Note that the tangent point to the 3--kpc arm is
at too high longitude for the model arm.
}
\endfigure

In Fig.\AILR (a), we show longitude--velocity trajectories
constructed by Mulder \& Liem (1986).
They solved gas--dynamical equations in a 
weakly--barred potential, using a finite--difference,
hydrodynamic grid code. The gas is described by the inviscid
Euler equations. 
The curves in Fig.\AILR (a) represent their preferred model
for the 3--kpc arm, the main feature used to constrain
the scales of their models. The viewing angle is
$\phi=$20\degr\ and $R_{\rm CR}=0.98$\rsun .
In Fig.\AILR (b), the model spiral is seen face--on.
Although this spiral reproduces the 3--kpc filament around
$\ell$=0\degr\ very well, the 
tangent point to the 3--kpc arm is at much too high longitude,
$\ell = -$27\degr\ (Fig.\AILR (a)).
Mulder \& Liem (1986) show the \lvd\ for another model, for
$\phi=$40\degr\ and $R_{\rm CR}=0.56$\rsun.
In this diagram the tangent point is much closer to the
observed point (see Fig.\HIS) and also the central
CO--disk's velocity signature is matched much better.
The trajectories of both ILR arms remain similar in the inner regions.
Unfortunately they do not give trajectories for this model but they
note that the arms are ``more concentrated around CR''.
This would be in agreement with our inner ring, discussed above.
The Mulder--Liem models
are forced to be quasi steady. Therefore, gas flows
around CR would not be represented correctly and it could 
well be that the two arms in Fig.\AILR (b) connect in reality
to form a ring. 

Binney \etal(1997) find, deprojecting the COBE map with
imposed eight--fold symmetry, two density enhancements on
the minor axis at 3 kpc from the Centre
($\ell = +17^{\circ},-22^{\circ}$). They suggest these
are the L$_{4,5}$ points. 
With four--fold symmetry, however, they find a,
leading, spiral. The longitude $\ell = -22^{\circ}$
of their feature coincides with the maxima in the OH/IR stars
and the 2.4 GHz emission used in this section.
As Binney \etal(1997) note, there may be a significant
contribution of young stars to the K--band surface density
(Rhoads 1996). We argue that their density features are 
incorrect deprojections of the inner ring.
In any case, the derived loci for CR are very similar.

The maxima \RRmu\ and \RRmc\ interestingly border those
of the 2.4 GHz emission (Fig.\SURF). Also, the locations of
the maxima in the older and younger OH/IR stars, respectively,
are slightly different. Such displacements may be caused 
by the intricate streaming and diffusion processes around
CR (Roberts, Huntley \& van Albada 1979; Kenney \& Lord 1991;
Vogel \etal1993; Tilanus \& Allen 1989), but it is 
virtually impossible to draw conclusions from them.

\subsection{The inner--Lindblad resonance}

Other than a corotation resonance and outer--Lindblad
resonances, the existence of an inner--Lindblad resonance (ILR)
in a galaxy is dependent on its exact potential.
If it exists, is may show via a a variety of indicators.
The ILR is usually outlined by
a "nuclear ring", often accompanied by massive star 
formation (see eg.~Buta 1996; Phillips 1996).  
A so--called
double--wave feature in the stellar rotation curve (Bettoni 1989)
may exist as a result of orbits trapped around the 
the retrograde $x_4$ orbit family inside ILR (Pfenniger 1984;
Wozniak \& Pfenniger 1997; Contopoulos \& Papayannopoulos 1980).
Finally, if an ILR exists, the gas flowing toward the centre,
following bar--induced instabilities, will follow offset spiral
arms (Athanassoula 1992). 
In absence of an ILR, the gas flows to the centre directly
along the major axis of the bar.

Comeron \& Torra (1996) 
report an elongated ring--shaped maximum in the deprojected 
density distribution of ultra--compact HII regions (UCHII).
They tentatively identify this with an ILR-- or nuclear ring
and give radii of 1.3 kpc and 1.9 kpc, for the 
southern-- and northern hemisphere, respectively
(corrected for \rsun=8 kpc). In general, these nuclear rings
are found to have radii around 1--1.5 kpc (Buta 1996; 
Freeman 1996).

The radio--continuum distribution (Fig.\RADIO) does
not show an increase in the number of small--scale sources 
at $\ell\sim-$\decdeg9.5 (1.3 kpc).
In that case, star formation must have started too recently for
massive stars to become supernovae ($\sim10^7$ yr) and
star formation must have ignited between
10$^7$ yr and 5$\times$10$^4$ yr ago (the ages of the UCHIIs).

A double--wave feature in the OH/IR--star rotation curve 
can indeed tentatively be identified in Fig.\VELO (a)
($\ell \sim +3^{\circ},-4^{\circ}$). 
As the AOSP sample is the only large--scale 
stellar--kinematical sample at low latitudes, it is 
impossible to verify this feature with other data.
According to Wozniak \& Pfenniger (1997), the 
deepest minimum in the rotation curve arises
around $0.25 R_{\rm CR}$, when the retrograde orbits
remain concentrated around the $x_4$ family.
It is not straightforward
to say where the maximum effect will take place in projection;
a simple tangent--point assumption is probably not valid.
The Max($E$)--model of Wozniak \& Pfenniger (1997, see their
Figs.4,5) is a good example.
In that model, the maximum effect, from the viewing point of the Sun,
of this ``counter--streaming'' comes from the minor 
axis of the bar. A location on the minor axis at $0.25 R_{\rm CR}$
would indeed be seen roughly at the longitudes of the 
observed double--wave feature.

In external barred galaxies, this feature is only seen at intermediate
inclinations (Bettoni \& Galletta 1997); the Galaxy might be the
only edge--on system that allows detection of the double wave,
as the disk--foreground contamination is relatively small with
respect to other galaxies. Some early--type galaxies
are known to harbour kinematically--distinct cores
(see eg. Carollo \etal1997) that can give a
similar feature in the rotation curve. The cores contribute
a much larger fraction to the total galaxy than do the
retrograde orbits mentioned above and so the feature is not
so easily masked by the foreground disk. We do not think
the feature observed in the Galaxy is caused by such a decoupled
core because the wave's velocity maximum and minimum are
small with respect to the rotation at large radii.
We checked that the feature is not the result of smearing
out the effect of the rapidly--rotating galactic--centre
OH/IR stars (see \S 3.2; Lindqvist \etal1992; Sjouwerman \etal1998a).

An offset gas flow around ILR would be virtually impossible to 
detect in the edge--on jungle of the galactic disk.
In Fig.\AILR (a), next to the 3--kpc stars, another
group of stars is highlighted.
The 13 stars forming the 
feature \RRmi\ (Fig.\SURF (e)), seen in the 
low--outflow, older OH/IR stars only, follow a
coherent filament in the \lvd\ that coincides with part
of the far ILR arm (see Fig.\AILR (b)).
All 13 stars lie on this filament, although in Fig.\AILR\ we have
plotted only 7 with $|b| < 0.6^{\circ}$. 
(Note that these stars were selected {\it before} overplotting them
on the ILR--arm model.)
In Fig.\AILR (b) we indicate roughly where the two small groups of
OH/IR stars would be located in the Mulder--Liem model (see
\S 5.1 for caveats concerning this model).
The compactness (see Fig.\SURF (e)) of this
group of relatively old stars, can indeed be understood more easily
if they are spread out along the \losn\ as shown.
However, the far--advanced age of this group of ILR--arm OH/IR stars
is puzzling and prevents us from accepting their location in this
picture too readily.

Exemplary of the ILR region in our Galaxy may be the inner region of the 
galaxy IC$\,$4214 (Hubble type SABab, Saraiva 1996).

\section{The present shape of the inner Galaxy}

\beginfigure{14}
\fignam\TOPV
{\psfig{figure=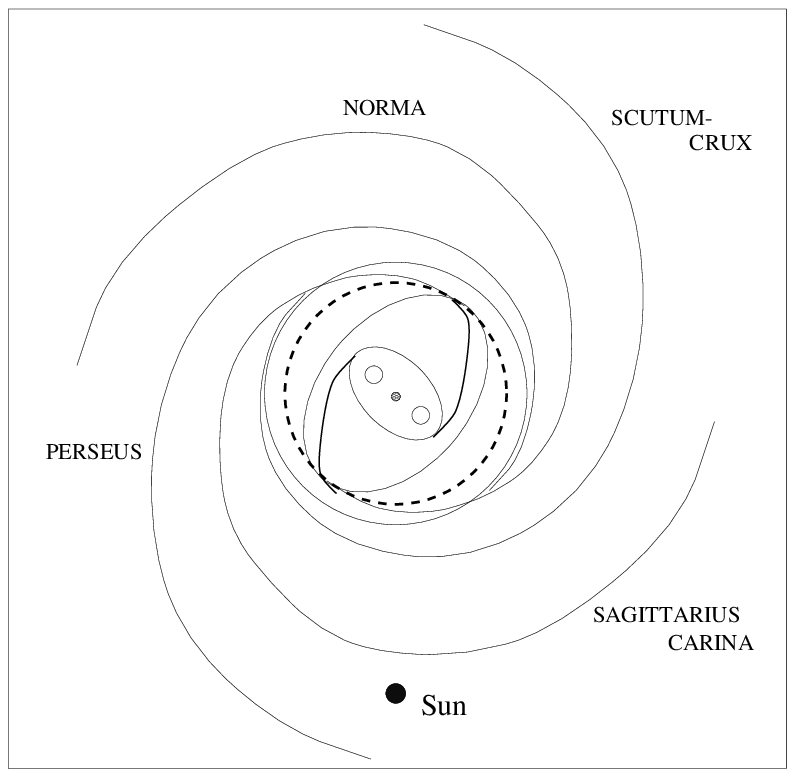}}
\caption{{\bf Figure \nfig}
Top--view sketch of the Galaxy as described in \S 6.
The larger ellipse outlines roughly the bar density; the
smaller the ILR. The two open circles inside the ILR indicate
the loci where the double--wave rotation feature may originate.
The short curves emerging from ILR are part of
the ILR arms (Mulder \& Liem 1986) connecting to IUHR--CR.
The thin, solid circle around the bar is CR. The thick
dashed ring indicates roughly the IUHR, the inner ring lies
between the IUHR-- and the CR circle.
The spiral arms outside CR are inspired
by the picture by Georgelin \& Georgelin (1976), structured 
by the assumption that the four spiral arms are two pairs,
bifurcating from one two--armed spiral inside CR.
The black dot marks the position of the Sun; the figure is
approximately to scale
with the distance of the Sun to the \gc\ (8 kpc).
Compare to Fig.6 in de Vaucouleurs \& Pence (1978), who sketch
a similar view of the Galaxy based on their derived
Hubble type (SA{\underbar B}({\underbar r}s)bc).
}
\endfigure

\noindent
The evidence presented in the previous sections
may be combined to give the following description
of the inner Galaxy (Fig.\TOPV).
There is a stellar Bar (\S 4.1) that is fairly flat 
and weak (\S 3;4.2). It corotates at $\sim$3.5 kpc and
is surrounded by an inner ring, inside corotation, 
roughly between 2.5 kpc and 3.5 kpc (\S 5.1). Possibly
an inner--Lindblad resonance exists, with corresponding nuclear
ring, between 1 kpc and 2 kpc (\S 5.2).
We suggest that outside corotation a 
four--armed, trailing spiral exists, 
which is a bifurcation of a two--armed spiral mode.
As such, the Galaxy would be similar in appearance to NGC$\,$1433 
(type (R$^{\prime}_1$)SB({\underbar r}s)ab; Buta 1995; for 
a picture see Buta 1996).

If we assume the rotation curve of the
axisymmetric part of the Galaxy is described by 184 \kmsr\ $R^{0.1}$
(Allen, Hyland \& Jones 1983),
a pattern speed of 60 \kmsr\ is implied for $R_{\rm CR}$=3.5 kpc.
The radii of ILR, IUHR and OLR in this case are 0.85 kpc, 2.2 kpc and
6.3 kpc, respectively. 
The radius of ILR depends
crucially on the adopted potential
and larger radii are easily possible in more centrally 
concentrated potentials.
The radius of IUHR, for 60 \kmsr , agrees well with our suggestion that the
inner ring, discussed in \S 5.1, lies between IUHR and CR.
A sudden decrease in metallicity is observed around $R=$ 6 kpc 
(Simpson \etal1995). If this marks the end of the zone--of--influence
of the Bar's mixing, as suggested by the authors,
one expects this to be around OLR. This agrees with the radius 
of OLR for a pattern speed of 60 \kmsr .

Comparison with published values for the corotation
radius and pattern speed proves little conclusive.
Values for corotation range from 
2.4 kpc (Binney \etal1991), 3.5 kpc (Weiner \& Sellwood 1996;
Englmaier \& Gerhard 1999), 
4.8 kpc (Fux 1997) to 9 kpc (Amaral \& Lepine 1997), 
and, similarly, for the pattern speed from 19\kmsr\ (Wada \etal1996)
to 118 \kmsr\ (Yuan 1984).

\subsection {Elongation}

In \S 4.2, we found that the kinematic type of 
the Bar indicates a rather weakly--barred distribution.
This is in agreement with the large 
in--plane axis ratio of the density, found in several studies :
$q_{\rho}\equiv q_{\Phi^{\prime \prime}}=0.6$ (Binney \etal1997, stellar 
density)
or  $q_{\rho}=0.7$ (Wada \etal1996, gas kinematics), although
Nikolaev \& Weinberg (1997) find $q_{\rho}=0.4$.
From Fig.\BARF , $q_{\rho} \sim 0.5$ is not excluded by our data.
Due to mixing (Friedli, Benz \& Kennicutt 1994), a correlation
between the in--plane axis ratio of bars and the metallicity
gradient of their host galaxies exists.
From an empirical relation between bar axis ratio and
[O/H] gradient (Martin \& Roy 1994), one finds
$q_{\rho}=0.7$ for a gradient of $-$0.07 dex kpc$^{-1}$ 
(Smartt \& Rolleston 1997).

Inner rings are usually elongated along the bar and follow the shape
of the main closed orbits; their axis ratio is $q_{\Phi^{\prime}}$ 
(see Buta 1996) and thus -- in general -- larger than $q_{\rho}$.
The elongation can be constrained by the radial motion along the ring.
At $\ell=$0\degr , the
\losa\ velocity is $-$53 \kms, or in other words
a radial velocity $V_{\rm R}$ of $+$53 \kms\ (Fig.\HIS; 
Oort 1977). If this is the 
\losa\ component of a streaming along the ring, $V_{\parallel}$,
we can define an angle $\beta= {\rm acos}(V_{\rm R}/V_{\parallel})$.
If we assume further that the ring is an ellipse, with a viewing 
angle $\phi$ of the major axis (near end at positive longitudes),
and that the streaming is along the ellipse, we can derive the
axis ratio $q_{\rm e}$ via $q_{\rm e}^2 = \rm tan(\beta-\phi)\,tan\,\phi$, with
$0 \le \phi < \beta \le {\pi \over 2}$.
For $V_{\parallel}=200$ \kms, $\beta=$ 75\degr\ and we find $q_{\rm e}=0.76$ for
$\phi=$ 45\degr\ ($q_{\rm e}=0.75$ for 
$\phi=$ 25\degr; $q_{\rm e}=0.61$ for $\phi=$ 65\degr).
For $V_{\parallel}=150-300$ \kms, $\beta=$ 70\degr--80\degr\ and 
$q_{\rm e}=0.68-0.84$. Hence, for $20^{\circ} < \phi < 55^{\circ}$,
the observed $V_{\rm R}$ is consistent with
$q_{\rm e}\equiv q_{\Phi^{\prime}} > 0.6$ for a wide range of $V_{\parallel}$.
The largest $q_{\Phi^{\prime}}$ arises for $\phi=0.5\, \beta$.

In general $q_{\Phi} > q_{\Phi^{\prime}} > q_{\rho}$, so the
radial velocity of the 3--kpc arm can be accommodated in an
almost--round potential (see also Sil'chenko \etal1997),
contrary to some claims (eg.~Mulder \& Liem 1986).

\section{Evolution of the inner Galaxy}

The morphological properties of bars are found to be related
to their formation mechanism (Noguchi 1996), to the
rotation curve (Combes \& Elmegreen 1993) and 
to the Hubble type of the host galaxy (Elmegreen \& Elmegreen 1985).
Early--type--galaxy bars (\lsim Sb in this context) have 
flat density profiles and an
ILR, end close to their corotation radius and form via strong tidal 
interaction with other galaxies. Late types (\gsim\ Sc)
have exponential density profiles, no ILR, end well inside 
(sometimes halfway) their
corotation radius and form via disk instabilities.

In this section, we will speculate on the formation and
evolution of the galactic Bar, and on how it influenced
the structure of the inner Galaxy.

\subsection{How did the Bar form ?}

The Galaxy is usually classified as an Sb or Sbc galaxy in the 
Hubble sequence (de Vaucouleurs \& Pence 1978)
and is as such just intermediate to the two classes described above.
The Bar's steep density profile (\S 3,4; 
Binney \etal1997) and
length ($R_{\rm CR}/a > R_{\rm CR}/R_{\rm IUHR} = 1.4$)
indicate a late type. Is it possible that the Bar in our
Galaxy indeed formed via disk instability$\,$?
Studies of Toomre's instability parameter
$Q$ (Toomre 1964) show that the Galaxy
is (marginally) stable 
at all radii (Fuchs \& von Linden 1997; Lewis \& Freeman 1989).
Earn \& Lynden--Bell (1996) also find that,
for the Bulge's present rotation curve
$V_{\rm c} \propto R^{0.1}$ ($\rho \propto R^{-1.8}$, Allen \etal1983),
$m=2$ modes will not be sustained via orbit cooperation (alignment).

However, in the first gigayears of the life of
the Galaxy ($\gg$5 Gyr ago), when the stellar dispersions
had not yet increased to their maximum value (Freeman 1991),
the disk was probably unstable (Fuchs \& von Linden 1997).
Especially if there was a weak trigger (the Large Magellanic Cloud,
Weinberg 1996) it is possible that the Bar formed
via a disk instability, more than 5 Gyr ago.
The strong correlation between kinematics and metallicity 
(eg. Zhao, Spergel \& Rich 1994)
favours secular bar formation and thus excludes strong
interaction as a mechanism, pointing to 
a disk--bar connection rather than 
to a halo--bar connection.
The similar scaleheights for the two components found in \S 3
(Table \HRZ) support a disk--bar connection and the formation
of the bar from the disk via a planar instability.

\subsection{When did the Bar form ?}

In general, it is suggested that the Bar is 
a young component with respect to the spheroidal components (Bulge, Halo).
Several authors find ages for the Bar of 6 Gyr to 9 Gyr
(Ng \etal1996; Gerhard \& Binney 1993; Wyse, Gilmore \& Franx 1997).
The OH/IR stars in the direction of the
Bar have an upper age limit of 7.5 Gyr (\S 2). If these OH/IR stars
are the result of renewed star formation in response to
the formation of the Bar (see discussion \S 7.3), 
then the Bar must be at least 8 Gyr old.
Such an age might allow for the disk--instability formation
mechanism (see above).
The scenario of Ng \etal(1996; their Fig.13) 
explains the absence of very old Bulge OH/IR stars
and the similarity of the metallicity of OH/IR stars
in the Bar and in the Disk (\S 2).
The absence of Bulge OH/IR stars older than 7.5 Gyr indicates that
the already existing old Bulge ($>$13 Gyr; Ng \etal1996; 
Wyse \etal1997) was not forming significant
numbers of 1--6 \msun\ stars in the plane
in this era.

\subsection{What happened next ?}

%The presence of a bar has a large influence on the 
%evolution of a galaxy.
Of special interest is the bar--induced mass flow to the
central regions of galaxies (Hasan, Pfenniger \& Norman 1993; Wada \& Habe
1992,1995) that has strong influence on the evolution of the
bar itself, eventually resulting in its destruction.
Wada \& Habe (1992, 1995) find that the presence
of an ILR is essential, but
however, Athanassoula (1992) and
Piner, Stone \& Teuben (1995) find that more generic
mass transport always occurs, even preferentially in systems without ILR. 

The gas, flowing to the centre,  causes instabilities and star formation.
In order to induce Jeans' instabilities at the scales of 
giant molecular clouds ($M_{\rm J} < M_{\rm GMC}$, 
eg.~Gerritsen \& Icke 1997) 
$10^8$\msun\ of gas has to be transported to within 100 pc 
(for $M_{\rm GMC}=1\times10^6$\msun; the masses of the SgrA,B GMCs
are $2,6\times10^6$\msun, Stark \etal1991).
The observed gas--inflow rate into the central 100 pc
is 0.01\msyr\ (von Linden, Duschl \& Biermann 1993) and
presumably was higher just after formation of the Bar 
(eg.~Piner \etal1995). This means it would take less than 10 Gyr,
or more or less the estimated age of the Bar,
for the central density to build up enough for the onset
of central star formation. 

Evidence for massive, on--going
star formation in the \gc\ from eg. the presence
of H$_2$O masers has been negated (Sjouwerman \& van Langevelde 1996).
Nevertheless, the rate inside 500 pc 
is higher than elsewhere, $\sim$0.5 \msyr\ (G\"usten
1989) and the
OH/IR stars in the \gc\ (inside 100 pc) appear to be younger
on average than the global population (Sjouwerman \etal1998b; 
Blommaert \etal1997; Wood, Habing \& McGregor 1998). 

The galactic--centre OH/IR stars have a 
bimodal distribution of outflows (Fig.\VEXP) that
differs from that of the Bulge sample 
with more than 99.9\% significance (Kolmogorov--Smirnov).
The lower outflow velocities are distributed in a rather
wide range (26\% fractional scatter) around 11.4 \kms, 
the higher in a narrow range (10\%) around 19.4 \kms .
Especially when allowing for a scatter in the metallicity,
the masses of the high--outflow 
galactic--centre stars must span a relatively
narrow range as well (Appendix A). This suggests a short episode of star 
formation, which according to Sjouwerman \etal(1998b) took place
more than 1 Gyr ago. The two groups of 
galactic--centre OH/IR stars are separated roughly by
\vexp = 15 \kms , corresponding to
an age of approximately 1.5 Gyr (\S 2, Appendix A)
for solar metallicity.
The sudden horizontal step in the cumulative distribution around
15 \kms\ suggests that around 1.5 Gyr ago there was 
either a sudden increase of the ambient metallicity or of 
the average mass of the formed stars.
Observations show that the high--outflow galactic--centre
OH/IR stars have metallicities of a factor of two higher
than the low--outflow-- and the Bulge objects (Blommaert \etal1997;
Wood \etal1998).
Hence, in the case of a burst starting 1.5 Gyr before present,
the progenitor masses of the present high--outflow population
must have been well over 3 \msun\ (Bertelli \etal1994) , in good
agreement with Wood \etal(1998). For twice--solar metallicity, 
the luminosity of these objects would be $\sim$90\% of the 
average luminosity of the Disk AOSP stars (Bertelli \etal1994) 
which, with the high outflow velocities, agrees well with equation (\MU).
Sjouwerman \etal(1998b) find that in total more than $10^7$\msun\ of 
stars would have formed in the starburst. However, they use 
half--solar metallicities in which case the luminosity would
be $\sim$110\% of the Disk AOSP stars (still for ages of 1.5 Gyr).
Such a model does not satisfy equation (\MU) at all. As 
this model uses lower progenitor masses, we conclude that 
in our high--metallicity scenario the burst would be a factor of five
times more massive than estimated by Sjouwerman \etal(1998b).

More recent ($\sim$ 7 Myr and $\sim$100 Myr ago)
and much less massive starbursts
are reported by Krabbe \etal(1995). These are too recent to 
show up in the outflow--velocity distribution of the OH/IR stars.

\subsection {The complete scenario}

The following speculative scenario emerges. 
The Bar formed $\sim$ 8 Gyr
ago via a disk instability.
Subsequent gas flows revived star formation in 
the Bulge region and the progenitors of the present--day
Bulge OH/IR stars started to form. The metallicity in the 
inner plane became similar for all radii. 
Enriched material kept flowing into the central 100 pc,
from the region inside corotation,
and around 1.5 Gyr ago, a star burst ignited,
leaving a significantly more metal--rich population
in the \gc .

Another possible consequence of the build--up of mass in the centre
is the formation of an ILR, typically when the mass inside 200 pc
is 1--2\% of the mass of the galaxy (Friedli \& Benz 1993). 
The ring of UCHIIs (\S 5.2) could indicate that an ILR 
exists in the Galaxy, and that has formed recently because it
is not seen in any older ($>$10 Myr) population.
The critical density for a nuclear--ring star burst
is $ 0.6 \kappa^2 /G$ (Elmegreen 1994a,b), yielding a required
ring mass of \lsim 1$\times10^8$ \msun\ which could have 
accumulated from the inflowing gas in $\sim$1 Gyr, once the ILR formed 
(0.1\msyr , Friedli \& Benz 1993, for $q_{\rho}=0.7$).
The ILR would hence have formed coevally with the central 
star burst derived from the OH/IR star population.

The total mass within 100 pc is $> 5\times10^8$\msun\
(Lindqvist \etal1992; Kent 1992).
This is $\sim$3\% of the total mass of the Bulge 
($\sim2\times10^{10}$ \msun; Zhao 1996; Blum 1995; Kent 1992).
Such a concentration is sufficient to 
destroy the Bar completely (Hasan \etal1993).

The ILR may slow down the self--destruction of the Bar,
by decreasing the rate of gas flow to the centre, 
but does not stop it (Hasan \etal1993; Piner \etal1995). 
In addition, inside ILR no bar--supporting orbits 
would exist (Contopoulos \& Papayannopoulos 1980).
With already an almost--critical amount of mass in the
central regions, the Bar may be completely dissolved
within a few Gyr.

A view of the possible future of the Galaxy can be obtained
by looking at NGC$\,$7217 (Athanassoula 1996).
This galaxy has an inner, an outer and a nuclear ring,
but no clear bar. It may be a remnant of
a once--barred galaxy that dissolved, which 
could also account for the high fraction (25\%) of
counter--rotating orbits in this galaxy (Merrifield \&
Kuijken, 1994).

The observed
CO parallelogram (Bally \etal1988) was elegantly explained
as the inner cusped $x_1$ orbit by Binney \etal(1991). 
However, Liszt \& Burton (1978) already showed that the
parallelogram is most likely the result of a tilt in the inner CO 
distribution. In this case, the constraint, placed on the
Bar's parameters by identifying this
inner cusped $x_1$ orbit (yielding 
a very low viewing angle and corotation radius), can be lifted.
In our scenario, the CO would be on $x_2$ orbits inside ILR.

\section{Conclusions}

We have given several new arguments in favour of the triaxiality of the
central Galaxy. 
In the inner 10\degr, perspective effects 
provide evidence for the existence of a Bar.
More indirect evidence comes from the presence of resonant features.
We argue that an inner ring, with a 
flat density distribution and rotation curve, and a gap in 
the Norma--arm region indicate the region inside corotation.
A possible nuclear ring and a double--wave feature in the stellar
rotation curve are indicative of the existence of an inner--Lindblad
resonance. We find that
corotation is at 3.5 kpc, the bar ends within 2.5 kpc (IUHR) and
the pattern speed is $\sim$ 60\kmsr .
The radius of the ILR is most likely larger than what would
be derived from a $R^{0.1}$ rotation curve (Allen \etal1983).

The much--studied 3--kpc arm is probably an inner ring,
connected to the IUHR--CR region 
rather than OLR (Yuan 1984; Binney \etal1991) or ILR 
(Amaral \& Lepine 1997).
Its radial motion at $\ell=$0\degr\ can be 
explained well by streaming along mildly 
elongated closed orbits ($q > 0.7$) for a wide
range of viewing angles.

The \gba\ is no extraordinary specimen.
The value of the anisotropy parameter $V_{\rm m} / \sigma_0$
and flattening $\epsilon$ define it as an SAB--bulge.
The Hubble type of the Galaxy
is SAB({\underbar r}s)b (and possibly an additional 
(nr) for the nuclear ring),
based on its axis ratio, its spiral and ring structure and
its kinematic type (cf. SA{\underbar B}({\underbar r}s)bc,
de Vaucouleurs \& Pence 1978).

Based on the maximum ages of the OH/IR stars 
in the direction of the Bulge, we speculate that the Bar's formation took
place $\sim$ 8 Gyr ago.
Induced mass flow to the centre changed the mass distribution
significantly. As a consequence, at $\sim$ 1.5 Gyr ago a
major star burst ignited (within 100 pc) and an ILR formed.
The Bar could be in the final stage of its existence.
%%%and has apparently undergone negligible vertical evolution.

The OH/IR stars in our sample
at low latitudes are all part of the intermediate--age
Bar and do not trace the old, axisymmetric Bulge 
mentioned by eg. Ng \etal(1996) and Wyse \etal(1997). 
The Bar determines, via its influence
on the metallicity gradient, the radial distribution of AGB stars
in the disk ($\sim$0.5 to 1.5 \rsun ), as
the ratio of oxygen--rich to carbon--rich AGB stars
is governed by the metallicity gradient. These two groups
together form one population in terms of their galactic distribution.
The AGB stars are distributed in the thin (old) disk with 
a scaleheight of 100 pc for the youngest AGB stars ( \lsim 1 Gyr)
and 500 pc for AGB stars older than \gsim 5 Gyr. 
There is possibly a disjunct 1--kpc--scaleheight population 
of OH/IR stars, that is seen only very locally ($\ell\sim$50\degr) and
is between 1 Gyr and 10 Gyr old (depending on metallicity).

\section*{Acknowledgments} 
I am grateful to the following people for their 
continued interest, their very helpful 
discussions, ideas, suggestions and for spotting easily--overlooked
mistakes: Richard Arnold, Frank van den Bosch, Butler Burton,
Carsten Dominik, Ron Ekers, Harm Habing, Vincent Icke,
Agris Kalnajs and Tim de Zeeuw.
Roger Fux made an N--body model available (Fig.\BARF)
and Lorant Sjouwerman OH/IR star data (Fig.\VEXP), both
before publication.
I also thank Janet Soulsby and Neil Killeen 
for helping improve my usage of the English language.

\vskip 1truecm

\appendixbegin {A} {The luminosity--metallicity--outflow--velocity relation}

\noindent
In this appendix, we outline the derivation of the relation between
(unknown) stellar luminosity, outflow velocity and envelope
metallicity for OH/IR stars.

We assume a star of luminosity $L_{\ast}$ and an isotropic
and constant
mass outflow $\dot M$. At a distance $R_0$ the gas temperature has
decreased to a value of $T_{\rm c}$, equal to the condensation 
temperature of some solid species. The particles that form are
immediately accelerated to their final outflow velocity \vexp.
Because of friction they take the gas along. We have now the following
equation :
\eqnam\AAA$$
  \dot M\, V_{\rm exp} = \tau L_{\ast}/c
\eqno(\new) $$
as the momentum change of the outflowing mass equals
the radiation pressure on the dust. $\tau$ is the dust opacity
of the \cse\ and $c$ the speed of light.
Also :
\eqnam\AAB$$
  \dot M = 4 \pi R_0^2 \rho_0 V_{\rm exp} 
\eqno(\new) $$
with $\rho_0$ the density at radius $R_0$.
From equation (\AAB) and the fact that $V_{\rm exp}$ is 
constant it follows that $\rho=\rho_0 (R_0/R)^2$.
We then obtain :
\eqnam\AAC$$
  \tau = {1 \over \mu} \kappa_0 \int_{R_0}^{\infty} \rho(R) {\rm d}R =
         {1 \over \mu} \kappa_0 \rho_0 R_0
\eqno(\new) $$
with $\kappa_0$ the mean absorption coefficient and $\mu$ the gas--to--dust
ratio of the \cse.
The dust forms when the radiation temperature has fallen to $T_{\rm c}$ :
\eqnam\AAD$$
   \sigma T^4_{\rm c} = L_{\ast}/(4 \pi R^2_0)
\eqno(\new) $$
with $\sigma$ the Stefan--Boltzmann constant.
We have four equations and seven variables ($\dot M, L_{\ast}, \mu, \tau,
R_0,\rho_0$,\vexp) and thus three parameters can be chosen freely.
We will take $\dot M, L_{\ast}$ and $\mu$. It is then easy to
show that :
\eqnam\AAE$$
   V_{\rm exp}^4 = A L_{\ast} \mu^{-2}
\eqno(\new) $$
where the constant A is given by 
\eqnam\AAF$$
    A = {\kappa_0^2 \sigma   \over (4\pi)^3 T_{\rm c}^4 c^2  }
\eqno(\new) $$
Thus \vexp\ is independent of $\dot M$. 

In a more elaborate model one may drop the assumption
that the dust acquires its final velocity immediately.
The gas is being dragged along by the dust; there will be
a drift between those components.
Analytic solutions no longer exist and equation (\AAE) is 
slightly different (Habing \etal1994) :
\eqnam\AAG$$
   V_{\rm exp}^{3.3} = A L_{\ast} \mu^{-1.7}
\eqno(\new) $$

\def\chaphead{B}
\beginfigure{15}
\fignam\IRS
\psfig{figure=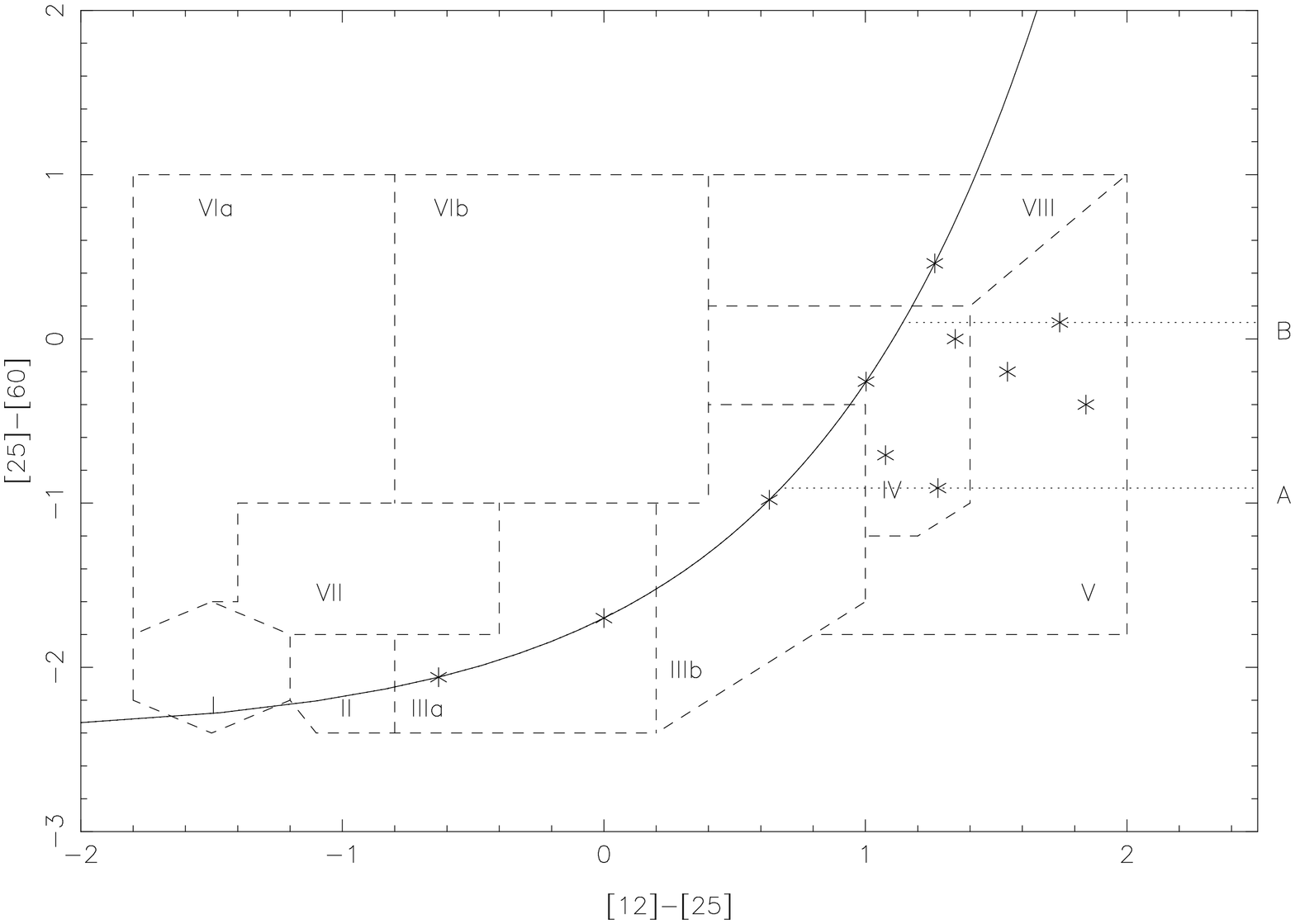,width=\hsize,angle=270}
\caption{{\bf Figure \nfig}
The IRAS two--colour diagram (van der Veen \& Habing 1988).
The colour $R_{21}$ equals
[12]-[25] which is defined as $ 2.5 \log S_{25\rm \mum}/S_{12\rm \mum} $
and accordingly $R_{32} \equiv $ [25]-[60].
The solid curve indicates the OH/IR--star
evolutionary track; the dashed
lines outline the various regions as they were defined by 
van der Veen \& Habing (1988). 
The two dotted lines give the largest and smallest $R_{32}^f$
(see text) for this fake sample.
}
\endfigure
\def\chaphead{A}

By dividing a sample of OH/IR stars into two according
to outflow velocity one thus effectively divides the sample
into more and less luminous stars. All existing models
of AGB--star evolution (eg. Vassiliades \& Wood) predict
that the more massive a main--sequence star is, the higher
its AGB luminosity. Thus stars of high \vexp\ are 
younger that those of low \vexp. This is confirmed by the
fact that stars with low \vexp\ have larger scaleheight
than those with high \vexp\ (first discussed by Baud \etal1981).

In principle there could be a conspiracy. The stars with 
higher \vexp\ could have higher metallicities and lower
values of $\mu$. A variation in $\mu$ could thus cancel
the effect of a variation in \vexp\ and stars with higher
\vexp\ would be {\it older}. However, for 
OH/IR stars, $\mu \propto Z^{-1}$ (Habing \etal1994).
For the conspiracy to work,  $Z \propto V^2_{\rm exp}$
and $Z$ must differ by at least a factor of four between stars
to explain the observed outflow velocities ($<$10 to $>$20 \kms).
It is therefore unlikely that the conspiracy should 
influence our conclusion, because the OH/IR stars are 
younger than $\sim$ 7 Gyr and oxygen--rich and thus have all 
formed in similar metallicity regimes (different by less
than a factor of three over 7 Gyr). A factor
of three difference in $Z$ may exist between the extremes
of the outer--Galaxy
and the metal--rich galactic--centre populations.

\vskip 1truecm

\appendixbegin {B} {Turn-off from the AGB--evolutionary track}

\noindent
In this appendix, we discuss a method to determine
upper age limits for an ensemble of OH/IR stars from 
their ``turn--over'' [25]-[60] colours in the IRAS two--colour
diagram.

According to Garcia Lario (1991), OH/IR stars evolve along
the evolutionary track (van der Veen \& Habing 1988) in the
IRAS two--colour diagram (Fig.\IRS) until they reach maximum
colours ($R_{32}^f,R_{21}^f$) dependent 
upon their main--sequence mass. At that
point, they leave the evolutionary track to evolve 
with approximately constant $R_{32} \equiv R_{32}^f$ toward higher $R_{21}$. 
The higher $R_{32}^f$, the higher the initial mass of a star :
\eqnam\ABA$$
  \log M_{\rm i} = \left({R_{32}^f + 2.42 \over 0.72 } - 2.45 \right)/3.2 .
\eqno(\new) $$
From Fig.\IRS\ it is seen how one can thus determine the upper and
lower limit to the initial masses and ages in a sample. The largest 
turn--over colour $R_{32}^f$ is indicated by B
and the smallest by A.  With equation (\ABA) we find the highest and
lowest masses present in the sample and with the isochrones
by Bertelli \etal(1994) we find the corresponding lowest
and highest ages. The lowest age thus found is an upper limit
to the true lowest age as there may be younger stars that have
not yet turned away from the evolutionary track (Fig.\IRS).

\bye

%% file: some_psfig.tex
% Psfig/TeX 
%%% Recupere sur /home/soft/.... a l'iap.
\def\PsfigVersion{1.9}
\ifx\undefined\psfig\else \fi

%
% from a suggestion by eijkhout@csrd.uiuc.edu to allow
% loading as a style file. Changed to avoid problems
% with amstex per suggestion by jbence@math.ucla.edu

\let\LaTeXAtSign=\@
\let\@=\relax
\edef\psfigRestoreAt{\catcode`\@=\number\catcode`@\relax}
\catcode`\@=11\relax
\newwrite\@unused
\def\ps@typeout#1{{\let\protect\string\immediate\write\@unused{#1}}}
\ps@typeout{psfig/tex \PsfigVersion}

%% Here's how you define your figure path.  Should be set up with null
%% default and a user useable definition.

\def\figurepath{./}

%
% @psdo control structure -- similar to Latex @for.
% I redefined these with different names so that psfig can
% be used with TeX as well as LaTeX, and so that it will not 
% be vunerable to future changes in LaTeX's internal
% control structure,
%
\def\@nnil{\@nil}
\def\@empty{}
\def\@psdonoop#1\@@#2#3{}
\def\@psdo#1:=#2\do#3{\edef\@psdotmp{#2}\ifx\@psdotmp\@empty \else
    \expandafter\@psdoloop#2,\@nil,\@nil\@@#1{#3}\fi}
\def\@psdoloop#1,#2,#3\@@#4#5{\def#4{#1}\ifx #4\@nnil \else
       #5\def#4{#2}\ifx #4\@nnil \else#5\@ipsdoloop #3\@@#4{#5}\fi\fi}
\def\@ipsdoloop#1,#2\@@#3#4{\def#3{#1}\ifx #3\@nnil 
       \let\@nextwhile=\@psdonoop \else
      #4\relax\let\@nextwhile=\@ipsdoloop\fi\@nextwhile#2\@@#3{#4}}
\def\@tpsdo#1:=#2\do#3{\xdef\@psdotmp{#2}\ifx\@psdotmp\@empty \else
    \@tpsdoloop#2\@nil\@nil\@@#1{#3}\fi}
\def\@tpsdoloop#1#2\@@#3#4{\def#3{#1}\ifx #3\@nnil 
       \let\@nextwhile=\@psdonoop \else
      #4\relax\let\@nextwhile=\@tpsdoloop\fi\@nextwhile#2\@@#3{#4}}
% 
% \fbox is defined in latex.tex; so if \fbox is undefined, assume that
% we are not in LaTeX.
% Perhaps this could be done better???
\ifx\undefined\fbox
% \fbox code from modified slightly from LaTeX
\newdimen\fboxrule
\newdimen\fboxsep
\newdimen\ps@tempdima
\newbox\ps@tempboxa
\fboxsep = 3pt
\fboxrule = .4pt
\long\def\fbox#1{\leavevmode\setbox\ps@tempboxa\hbox{#1}\ps@tempdima\fboxrule
    \advance\ps@tempdima \fboxsep \advance\ps@tempdima \dp\ps@tempboxa
   \hbox{\lower \ps@tempdima\hbox
  {\vbox{\hrule height \fboxrule
          \hbox{\vrule width \fboxrule \hskip\fboxsep
          \vbox{\vskip\fboxsep \box\ps@tempboxa\vskip\fboxsep}\hskip 
                 \fboxsep\vrule width \fboxrule}
                 \hrule height \fboxrule}}}}
\fi
%
%%%%%%%%%%%%%%%%%%%%%%%%%%%%%%%%%%%%%%%%%%%%%%%%%%%%%%%%%%%%%%%%%%%
% file reading stuff from epsf.tex
%   EPSF.TEX macro file:
%   Written by Tomas Rokicki of Radical Eye Software, 29 Mar 1989.
%   Revised by Don Knuth, 3 Jan 1990.
%   Revised by Tomas Rokicki to accept bounding boxes with no
%      space after the colon, 18 Jul 1990.
%   Portions modified/removed for use in PSFIG package by
%      J. Daniel Smith, 9 October 1990.
%
\newread\ps@stream
\newif\ifnot@eof       % continue looking for the bounding box?
\newif\if@noisy        % report what you're making?
\newif\if@atend        % %%BoundingBox: has (at end) specification
\newif\if@psfile       % does this look like a PostScript file?
%
% PostScript files should start with `%!'
%
{\catcode`\%=12\global\gdef\epsf@start{%!}}
\def\epsf@PS{PS}
\def\epsf@getbb#1{%
%
%   The first thing we need to do is to open the
%   PostScript file, if possible.
%
\openin\ps@stream=#1
\ifeof\ps@stream\ps@typeout{Error, File #1 not found}\else
%
%   Okay, we got it. Now we'll scan lines until we find one that doesn't
%   start with %. We're looking for the bounding box comment.
%
   {\not@eoftrue \chardef\other=12
    \def\do##1{\catcode`##1=\other}\dospecials \catcode`\ =10
    \loop
       \if@psfile
	  \read\ps@stream to \epsf@fileline
       \else{
	  \obeyspaces
          \read\ps@stream to \epsf@tmp\global\let\epsf@fileline\epsf@tmp}
       \fi
       \ifeof\ps@stream\not@eoffalse\else
%
%   Check the first line for `%!'.  Issue a warning message if its not
%   there, since the file might not be a PostScript file.
%
       \if@psfile\else
       \expandafter\epsf@test\epsf@fileline:. \\%
       \fi
%
%   We check to see if the first character is a % sign;
%   if so, we look further and stop only if the line begins with
%   `%%BoundingBox:' and the `(atend)' specification was not found.
%   That is, the only way to stop is when the end of file is reached,
%   or a `%%BoundingBox: llx lly urx ury' line is found.
%
          \expandafter\epsf@aux\epsf@fileline:. \\%
       \fi
   \ifnot@eof\repeat
   }\closein\ps@stream\fi}%
%
% This tests if the file we are reading looks like a PostScript file.
%
\long\def\epsf@test#1#2#3:#4\\{\def\epsf@testit{#1#2}
			\ifx\epsf@testit\epsf@start\else
\ps@typeout{Warning! File does not start with `\epsf@start'.  It may not be a PostScript file.}
			\fi
			\@psfiletrue} % don't test after 1st line
%
%   We still need to define the tricky \epsf@aux macro. This requires
%   a couple of magic constants for comparison purposes.
%
{\catcode`\%=12\global\let\epsf@percent=%\global\def\epsf@bblit{%BoundingBox}}
%
%
%   So we're ready to check for `%BoundingBox:' and to grab the
%   values if they are found.  We continue searching if `(at end)'
%   was found after the `%BoundingBox:'.
%
\long\def\epsf@aux#1#2:#3\\{\ifx#1\epsf@percent
   \def\epsf@testit{#2}\ifx\epsf@testit\epsf@bblit
	\@atendfalse
        \epsf@atend #3 . \\%
	\if@atend	
	   \if@verbose{
		\ps@typeout{psfig: found `(atend)'; continuing search}
	   }\fi
        \else
        \epsf@grab #3 . . . \\%
        \not@eoffalse
        \global\no@bbfalse
        \fi
   \fi\fi}%
%
%   Here we grab the values and stuff them in the appropriate definitions.
%
\def\epsf@grab #1 #2 #3 #4 #5\\{%
   \global\def\epsf@llx{#1}\ifx\epsf@llx\empty
      \epsf@grab #2 #3 #4 #5 .\\\else
   \global\def\epsf@lly{#2}%
   \global\def\epsf@urx{#3}\global\def\epsf@ury{#4}\fi}%
%
% Determine if the stuff following the %%BoundingBox is `(atend)'
% J. Daniel Smith.  Copied from \epsf@grab above.
%
\def\epsf@atendlit{(atend)} 
\def\epsf@atend #1 #2 #3\\{%
   \def\epsf@tmp{#1}\ifx\epsf@tmp\empty
      \epsf@atend #2 #3 .\\\else
   \ifx\epsf@tmp\epsf@atendlit\@atendtrue\fi\fi}

% End of file reading stuff from epsf.tex
%%%%%%%%%%%%%%%%%%%%%%%%%%%%%%%%%%%%%%%%%%%%%%%%%%%%%%%%%%%%%%%%%%%

%%%%%%%%%%%%%%%%%%%%%%%%%%%%%%%%%%%%%%%%%%%%%%%%%%%%%%%%%%%%%%%%%%%
% trigonometry stuff from "trig.tex"
\chardef\psletter = 11 % won't conflict with \begin{letter} now...
\chardef\other = 12

\newif \ifdebug %%% turn me on to see TeX hard at work ...
\newif\ifc@mpute %%% don't need to compute some values
\c@mputetrue % but assume that we do

\let\then = \relax
\def\r@dian{pt }
\let\r@dians = \r@dian
\let\dimensionless@nit = \r@dian
\let\dimensionless@nits = \dimensionless@nit
\def\internal@nit{sp }
\let\internal@nits = \internal@nit
\newif\ifstillc@nverging
\def \Mess@ge #1{\ifdebug \then \message {#1} \fi}

{ %%% Things that need abnormal catcodes %%%
	\catcode `\@ = \psletter
	\gdef \nodimen {\expandafter \n@dimen \the \dimen}
	\gdef \term #1 #2 #3%
	       {\edef \t@ {\the #1}%%% freeze parameter 1 (count, by value)
		\edef \t@@ {\expandafter \n@dimen \the #2\r@dian}%
				   %%% freeze parameter 2 (dimen, by value)
		\t@rm {\t@} {\t@@} {#3}%
	       }
	\gdef \t@rm #1 #2 #3%
	       {{%
		\count 0 = 0
		\dimen 0 = 1 \dimensionless@nit
		\dimen 2 = #2\relax
		\Mess@ge {Calculating term #1 of \nodimen 2}%
		\loop
		\ifnum	\count 0 < #1
		\then	\advance \count 0 by 1
			\Mess@ge {Iteration \the \count 0 \space}%
			\Multiply \dimen 0 by {\dimen 2}%
			\Mess@ge {After multiplication, term = \nodimen 0}%
			\Divide \dimen 0 by {\count 0}%
			\Mess@ge {After division, term = \nodimen 0}%
		\repeat
		\Mess@ge {Final value for term #1 of 
				\nodimen 2 \space is \nodimen 0}%
		\xdef \Term {#3 = \nodimen 0 \r@dians}%
		\aftergroup \Term
	       }}
	\catcode `\p = \other
	\catcode `\t = \other
	\gdef \n@dimen #1pt{#1} %%% throw away the ``pt''
}

\def \Divide #1by #2{\divide #1 by #2} %%% just a synonym

\def \Multiply #1by #2%%% allows division of a dimen by a dimen
       {{%%% should really freeze parameter 2 (dimen, passed by value)
	\count 0 = #1\relax
	\count 2 = #2\relax
	\count 4 = 65536
	\Mess@ge {Before scaling, count 0 = \the \count 0 \space and
			count 2 = \the \count 2}%
	\ifnum	\count 0 > 32767 %%% do our best to avoid overflow
	\then	\divide \count 0 by 4
		\divide \count 4 by 4
	\else	\ifnum	\count 0 < -32767
		\then	\divide \count 0 by 4
			\divide \count 4 by 4
		\else
		\fi
	\fi
	\ifnum	\count 2 > 32767 %%% while retaining reasonable accuracy
	\then	\divide \count 2 by 4
		\divide \count 4 by 4
	\else	\ifnum	\count 2 < -32767
		\then	\divide \count 2 by 4
			\divide \count 4 by 4
		\else
		\fi
	\fi
	\multiply \count 0 by \count 2
	\divide \count 0 by \count 4
	\xdef \product {#1 = \the \count 0 \internal@nits}%
	\aftergroup \product
       }}

\def\r@duce{\ifdim\dimen0 > 90\r@dian \then   % sin(x+90) = sin(180-x)
		\multiply\dimen0 by -1
		\advance\dimen0 by 180\r@dian
		\r@duce
	    \else \ifdim\dimen0 < -90\r@dian \then  % sin(-x) = sin(360+x)
		\advance\dimen0 by 360\r@dian
		\r@duce
		\fi
	    \fi}

\def\Sine#1%
       {{%
	\dimen 0 = #1 \r@dian
	\r@duce
	\ifdim\dimen0 = -90\r@dian \then
	   \dimen4 = -1\r@dian
	   \c@mputefalse
	\fi
	\ifdim\dimen0 = 90\r@dian \then
	   \dimen4 = 1\r@dian
	   \c@mputefalse
	\fi
	\ifdim\dimen0 = 0\r@dian \then
	   \dimen4 = 0\r@dian
	   \c@mputefalse
	\fi
	\ifc@mpute \then
        	% convert degrees to radians
		\divide\dimen0 by 180
		\dimen0=3.141592654\dimen0
		\dimen 2 = 3.1415926535897963\r@dian %%% a well-known constant
		\divide\dimen 2 by 2 %%% we only deal with -pi/2 : pi/2
		\Mess@ge {Sin: calculating Sin of \nodimen 0}%
		\count 0 = 1 %%% see power-series expansion for sine
		\dimen 2 = 1 \r@dian %%% ditto
		\dimen 4 = 0 \r@dian %%% ditto
		\loop
			\ifnum	\dimen 2 = 0 %%% then we've done
			\then	\stillc@nvergingfalse 
			\else	\stillc@nvergingtrue
			\fi
			\ifstillc@nverging %%% then calculate next term
			\then	\term {\count 0} {\dimen 0} {\dimen 2}%
				\advance \count 0 by 2
				\count 2 = \count 0
				\divide \count 2 by 2
				\ifodd	\count 2 %%% signs alternate
				\then	\advance \dimen 4 by \dimen 2
				\else	\advance \dimen 4 by -\dimen 2
				\fi
		\repeat
	\fi		
			\xdef \sine {\nodimen 4}%
       }}

% Now the Cosine can be calculated easily by calling \Sine
\def\Cosine#1{\ifx\sine\UnDefined\edef\Savesine{\relax}\else
		             \edef\Savesine{\sine}\fi
	{\dimen0=#1\r@dian\advance\dimen0 by 90\r@dian
	 \Sine{\nodimen 0}
	 \xdef\cosine{\sine}
	 \xdef\sine{\Savesine}}}	      
% end of trig stuff
%%%%%%%%%%%%%%%%%%%%%%%%%%%%%%%%%%%%%%%%%%%%%%%%%%%%%%%%%%%%%%%%%%%%

\def\psdraft{
	\def\@psdraft{0}
	%\ps@typeout{draft level now is \@psdraft \space . }
}
\def\psfull{
	\def\@psdraft{100}
	%\ps@typeout{draft level now is \@psdraft \space . }
}

\psfull

\newif\if@scalefirst
\def\psscalefirst{\@scalefirsttrue}
\def\psrotatefirst{\@scalefirstfalse}
\psrotatefirst

\newif\if@draftbox
\def\psnodraftbox{
	\@draftboxfalse
}
\def\psdraftbox{
	\@draftboxtrue
}
\@draftboxtrue

\newif\if@prologfile
\newif\if@postlogfile
\def\pssilent{
	\@noisyfalse
}
\def\psnoisy{
	\@noisytrue
}
\psnoisy
%%% These are for the option list.
%%% A specification of the form a = b maps to calling \@p@@sa{b}
\newif\if@bbllx
\newif\if@bblly
\newif\if@bburx
\newif\if@bbury
\newif\if@height
\newif\if@width
\newif\if@rheight
\newif\if@rwidth
\newif\if@angle
\newif\if@clip
\newif\if@verbose
\def\@p@@sclip#1{\@cliptrue}

\newif\if@decmpr

%%% GDH 7/26/87 -- changed so that it first looks in the local directory,
%%% then in a specified global directory for the ps file.
%%% RPR 6/25/91 -- changed so that it defaults to user-supplied name if
%%% boundingbox info is specified, assuming graphic will be created by
%%% print time.
%%% TJD 10/19/91 -- added bbfile vs. file distinction, and @decmpr flag

\def\@p@@sfigure#1{\def\@p@sfile{null}\def\@p@sbbfile{null}
	        \openin1=#1.bb
		\ifeof1\closein1
	        	\openin1=\figurepath#1.bb
			\ifeof1\closein1
			        \openin1=#1
				\ifeof1\closein1%
				       \openin1=\figurepath#1
					\ifeof1
					   \ps@typeout{Error, File #1 not found}
						\if@bbllx\if@bblly
				   		\if@bburx\if@bbury
			      				\def\@p@sfile{#1}%
			      				\def\@p@sbbfile{#1}%
							\@decmprfalse
				  	   	\fi\fi\fi\fi
					\else\closein1
				    		\def\@p@sfile{\figurepath#1}%
				    		\def\@p@sbbfile{\figurepath#1}%
						\@decmprfalse
	                       		\fi%
			 	\else\closein1%
					\def\@p@sfile{#1}
					\def\@p@sbbfile{#1}
					\@decmprfalse
			 	\fi
			\else
				\def\@p@sfile{\figurepath#1}
				\def\@p@sbbfile{\figurepath#1.bb}
				\@decmprtrue
			\fi
		\else
			\def\@p@sfile{#1}
			\def\@p@sbbfile{#1.bb}
			\@decmprtrue
		\fi}

\def\@p@@sfile#1{\@p@@sfigure{#1}}

\def\@p@@sbbllx#1{
		%\ps@typeout{bbllx is #1}
		\@bbllxtrue
		\dimen100=#1
		\edef\@p@sbbllx{\number\dimen100}
}
\def\@p@@sbblly#1{
		%\ps@typeout{bblly is #1}
		\@bbllytrue
		\dimen100=#1
		\edef\@p@sbblly{\number\dimen100}
}
\def\@p@@sbburx#1{
		%\ps@typeout{bburx is #1}
		\@bburxtrue
		\dimen100=#1
		\edef\@p@sbburx{\number\dimen100}
}
\def\@p@@sbbury#1{
		%\ps@typeout{bbury is #1}
		\@bburytrue
		\dimen100=#1
		\edef\@p@sbbury{\number\dimen100}
}
\def\@p@@sheight#1{
		\@heighttrue
		\dimen100=#1
   		\edef\@p@sheight{\number\dimen100}
		%\ps@typeout{Height is \@p@sheight}
}
\def\@p@@swidth#1{
		%\ps@typeout{Width is #1}
		\@widthtrue
		\dimen100=#1
		\edef\@p@swidth{\number\dimen100}
}
\def\@p@@srheight#1{
		%\ps@typeout{Reserved height is #1}
		\@rheighttrue
		\dimen100=#1
		\edef\@p@srheight{\number\dimen100}
}
\def\@p@@srwidth#1{
		%\ps@typeout{Reserved width is #1}
		\@rwidthtrue
		\dimen100=#1
		\edef\@p@srwidth{\number\dimen100}
}
\def\@p@@sangle#1{
		%\ps@typeout{Rotation is #1}
		\@angletrue
%		\dimen100=#1
		\edef\@p@sangle{#1} %\number\dimen100}
}
\def\@p@@ssilent#1{ 
		\@verbosefalse
}
\def\@p@@sprolog#1{\@prologfiletrue\def\@prologfileval{#1}}
\def\@p@@spostlog#1{\@postlogfiletrue\def\@postlogfileval{#1}}
\def\@cs@name#1{\csname #1\endcsname}
\def\@setparms#1=#2,{\@cs@name{@p@@s#1}{#2}}
%
% initialize the defaults (size the size of the figure)
%
\def\ps@init@parms{
		\@bbllxfalse \@bbllyfalse
		\@bburxfalse \@bburyfalse
		\@heightfalse \@widthfalse
		\@rheightfalse \@rwidthfalse
		\def\@p@sbbllx{}\def\@p@sbblly{}
		\def\@p@sbburx{}\def\@p@sbbury{}
		\def\@p@sheight{}\def\@p@swidth{}
		\def\@p@srheight{}\def\@p@srwidth{}
		\def\@p@sangle{0}
		\def\@p@sfile{} \def\@p@sbbfile{}
		\def\@p@scost{10}
		\def\@sc{}
		\@prologfilefalse
		\@postlogfilefalse
		\@clipfalse
		\if@noisy
			\@verbosetrue
		\else
			\@verbosefalse
		\fi
}
%
% Go through the options setting things up.
%
\def\parse@ps@parms#1{
	 	\@psdo\@psfiga:=#1\do
		   {\expandafter\@setparms\@psfiga,}}
%
% Compute bb height and width
%
\newif\ifno@bb
\def\bb@missing{
	\if@verbose{
		\ps@typeout{psfig: searching \@p@sbbfile \space  for bounding box}
	}\fi
	\no@bbtrue
	\epsf@getbb{\@p@sbbfile}
        \ifno@bb \else \bb@cull\epsf@llx\epsf@lly\epsf@urx\epsf@ury\fi
}	
\def\bb@cull#1#2#3#4{
	\dimen100=#1 bp\edef\@p@sbbllx{\number\dimen100}
	\dimen100=#2 bp\edef\@p@sbblly{\number\dimen100}
	\dimen100=#3 bp\edef\@p@sbburx{\number\dimen100}
	\dimen100=#4 bp\edef\@p@sbbury{\number\dimen100}
	\no@bbfalse
}
% rotate point (#1,#2) about (0,0).
% The sine and cosine of the angle are already stored in \sine and
% \cosine.  The result is placed in (\p@intvaluex, \p@intvaluey).
\newdimen\p@intvaluex
\newdimen\p@intvaluey
\def\rotate@#1#2{{\dimen0=#1 sp\dimen1=#2 sp
%            	calculate x' = x \cos\theta - y \sin\theta
		  \global\p@intvaluex=\cosine\dimen0
		  \dimen3=\sine\dimen1
		  \global\advance\p@intvaluex by -\dimen3
% 		calculate y' = x \sin\theta + y \cos\theta
		  \global\p@intvaluey=\sine\dimen0
		  \dimen3=\cosine\dimen1
		  \global\advance\p@intvaluey by \dimen3
		  }}
\def\compute@bb{
		\no@bbfalse
		\if@bbllx \else \no@bbtrue \fi
		\if@bblly \else \no@bbtrue \fi
		\if@bburx \else \no@bbtrue \fi
		\if@bbury \else \no@bbtrue \fi
		\ifno@bb \bb@missing \fi
		\ifno@bb \ps@typeout{FATAL ERROR: no bb supplied or found}
			\no-bb-error
		\fi
		%
%\ps@typeout{BB: \@p@sbbllx, \@p@sbblly, \@p@sbburx, \@p@sbbury} 
%
% store height/width of original (unrotated) bounding box
		\count203=\@p@sbburx
		\count204=\@p@sbbury
		\advance\count203 by -\@p@sbbllx
		\advance\count204 by -\@p@sbblly
		\edef\ps@bbw{\number\count203}
		\edef\ps@bbh{\number\count204}
		%\ps@typeout{ psbbh = \ps@bbh, psbbw = \ps@bbw }
		\if@angle 
			\Sine{\@p@sangle}\Cosine{\@p@sangle}
	        	{\dimen100=\maxdimen\xdef\r@p@sbbllx{\number\dimen100}
					    \xdef\r@p@sbblly{\number\dimen100}
			                    \xdef\r@p@sbburx{-\number\dimen100}
					    \xdef\r@p@sbbury{-\number\dimen100}}
%
% Need to rotate all four points and take the X-Y extremes of the new
% points as the new bounding box.
                        \def\minmaxtest{
			   \ifnum\number\p@intvaluex<\r@p@sbbllx
			      \xdef\r@p@sbbllx{\number\p@intvaluex}\fi
			   \ifnum\number\p@intvaluex>\r@p@sbburx
			      \xdef\r@p@sbburx{\number\p@intvaluex}\fi
			   \ifnum\number\p@intvaluey<\r@p@sbblly
			      \xdef\r@p@sbblly{\number\p@intvaluey}\fi
			   \ifnum\number\p@intvaluey>\r@p@sbbury
			      \xdef\r@p@sbbury{\number\p@intvaluey}\fi
			   }
%			lower left
			\rotate@{\@p@sbbllx}{\@p@sbblly}
			\minmaxtest
%			upper left
			\rotate@{\@p@sbbllx}{\@p@sbbury}
			\minmaxtest
%			lower right
			\rotate@{\@p@sbburx}{\@p@sbblly}
			\minmaxtest
%			upper right
			\rotate@{\@p@sbburx}{\@p@sbbury}
			\minmaxtest
			\edef\@p@sbbllx{\r@p@sbbllx}\edef\@p@sbblly{\r@p@sbblly}
			\edef\@p@sbburx{\r@p@sbburx}\edef\@p@sbbury{\r@p@sbbury}
%\ps@typeout{rotated BB: \r@p@sbbllx, \r@p@sbblly, \r@p@sbburx, \r@p@sbbury}
		\fi
		\count203=\@p@sbburx
		\count204=\@p@sbbury
		\advance\count203 by -\@p@sbbllx
		\advance\count204 by -\@p@sbblly
		\edef\@bbw{\number\count203}
		\edef\@bbh{\number\count204}
		%\ps@typeout{ bbh = \@bbh, bbw = \@bbw }
}
%
% \in@hundreds performs #1 * (#2 / #3) correct to the hundreds,
%	then leaves the result in @result
%
\def\in@hundreds#1#2#3{\count240=#2 \count241=#3
		     \count100=\count240	% 100 is first digit #2/#3
		     \divide\count100 by \count241
		     \count101=\count100
		     \multiply\count101 by \count241
		     \advance\count240 by -\count101
		     \multiply\count240 by 10
		     \count101=\count240	%101 is second digit of #2/#3
		     \divide\count101 by \count241
		     \count102=\count101
		     \multiply\count102 by \count241
		     \advance\count240 by -\count102
		     \multiply\count240 by 10
		     \count102=\count240	% 102 is the third digit
		     \divide\count102 by \count241
		     \count200=#1\count205=0
		     \count201=\count200
			\multiply\count201 by \count100
		 	\advance\count205 by \count201
		     \count201=\count200
			\divide\count201 by 10
			\multiply\count201 by \count101
			\advance\count205 by \count201
		     \count201=\count200
			\divide\count201 by 100
			\multiply\count201 by \count102
			\advance\count205 by \count201
		     \edef\@result{\number\count205}
}
\def\compute@wfromh{
		% computing : width = height * (bbw / bbh)
		\in@hundreds{\@p@sheight}{\@bbw}{\@bbh}
		%\ps@typeout{ \@p@sheight * \@bbw / \@bbh, = \@result }
		\edef\@p@swidth{\@result}
		%\ps@typeout{w from h: width is \@p@swidth}
}
\def\compute@hfromw{
		% computing : height = width * (bbh / bbw)
	        \in@hundreds{\@p@swidth}{\@bbh}{\@bbw}
		%\ps@typeout{ \@p@swidth * \@bbh / \@bbw = \@result }
		\edef\@p@sheight{\@result}
		%\ps@typeout{h from w : height is \@p@sheight}
}
\def\compute@handw{
		\if@height 
			\if@width
			\else
				\compute@wfromh
			\fi
		\else 
			\if@width
				\compute@hfromw
			\else
				\edef\@p@sheight{\@bbh}
				\edef\@p@swidth{\@bbw}
			\fi
		\fi
}
\def\compute@resv{
		\if@rheight \else \edef\@p@srheight{\@p@sheight} \fi
		\if@rwidth \else \edef\@p@srwidth{\@p@swidth} \fi
		%\ps@typeout{rheight = \@p@srheight, rwidth = \@p@srwidth}
}
%		
% Compute any missing values
\def\compute@sizes{
	\compute@bb
	\if@scalefirst\if@angle
% at this point the bounding box has been adjsuted correctly for
% rotation.  PSFIG does all of its scaling using \@bbh and \@bbw.  If
% a width= or height= was specified along with \psscalefirst, then the
% width=/height= value needs to be adjusted to match the new (rotated)
% bounding box size (specifed in \@bbw and \@bbh).
%    \ps@bbw       width=
%    -------  =  ---------- 
%    \@bbw       new width=
% so `new width=' = (width= * \@bbw) / \ps@bbw; where \ps@bbw is the
% width of the original (unrotated) bounding box.
	\if@width
	   \in@hundreds{\@p@swidth}{\@bbw}{\ps@bbw}
	   \edef\@p@swidth{\@result}
	\fi
	\if@height
	   \in@hundreds{\@p@sheight}{\@bbh}{\ps@bbh}
	   \edef\@p@sheight{\@result}
	\fi
	\fi\fi
	\compute@handw
	\compute@resv}

%
% \psfig
% usage : \psfig{file=, height=, width=, bbllx=, bblly=, bburx=, bbury=,
%			rheight=, rwidth=, clip=}
%
% "clip=" is a switch and takes no value, but the `=' must be present.
\def\psfig#1{\vbox {
	% do a zero width hard space so that a single
	% \psfig in a centering enviornment will behave nicely
	%{\setbox0=\hbox{\ }\ \hskip-\wd0}
	%
	\ps@init@parms
	\parse@ps@parms{#1}
	\compute@sizes
	\ifnum\@p@scost<\@psdraft{
		\special{ps::[begin] 	\@p@swidth \space \@p@sheight \space
				\@p@sbbllx \space \@p@sbblly \space
				\@p@sbburx \space \@p@sbbury \space
				startTexFig \space }
		\if@angle
			\special {ps:: \@p@sangle \space rotate \space} 
		\fi
		\if@clip{
			\if@verbose{
				\ps@typeout{(clip)}
			}\fi
			\special{ps:: doclip \space }
		}\fi
		\if@prologfile
		    \special{ps: plotfile \@prologfileval \space } \fi
		\if@decmpr{
			\if@verbose{
				\ps@typeout{psfig: including \@p@sfile.Z \space }
			}\fi
			\special{ps: plotfile "`zcat \@p@sfile.Z" \space }
		}\else{
			\if@verbose{
				\ps@typeout{psfig: including \@p@sfile \space }
			}\fi
			\special{ps: plotfile \@p@sfile \space }
		}\fi
		\if@postlogfile
		    \special{ps: plotfile \@postlogfileval \space } \fi
		\special{ps::[end] endTexFig \space }
		% Create the vbox to reserve the space for the figure.
		\vbox to \@p@srheight sp{
		% 1/92 TJD Changed from "true sp" to "sp" for magnification.
			\hbox to \@p@srwidth sp{
				\hss
			}
		\vss
		}
	}\else{
		% draft figure, just reserve the space and print the
		% path name.
		\if@draftbox{		
			% Verbose draft: print file name in box
			\hbox{\frame{\vbox to \@p@srheight sp{
			\vss
			\hbox to \@p@srwidth sp{ \hss \@p@sfile \hss }
			\vss
			}}}
		}\else{
			% Non-verbose draft
			\vbox to \@p@srheight sp{
			\vss
			\hbox to \@p@srwidth sp{\hss}
			\vss
			}
		}\fi

	}\fi
}}
\psfigRestoreAt
\let\@=\LaTeXAtSign

%% file: some_incl.tex
%%%%%%%%%%%%%%%%%%%%%%%%%%%%%%%%%%%%%%%%%%%%%%%%%%%%%%%%%%%%%%%%%%%%%%%%%%%%%%
%
%                         Intro 
%
%%%%%%%%%%%%%%%%%%%%%%%%%%%%%%%%%%%%%%%%%%%%%%%%%%%%%%%%%%%%%%%%%%%%%%%%%%%%%%
\def\bck{\hskip-0.35em}
\def\wisk#1{\ifmmode{#1}\else{$#1$}\fi} 
\def\extra#1{\wisk{\phantom{\rm#1}}}
\def\gt   {$\!$\hbox{\tt >}$\!$}
\def\lt   {$\!$\hbox{\tt <}$\!$}
\def\oversim#1#2{\lower1.5pt\vbox{\baselineskip0pt \lineskip-0.5pt
     \ialign{$\mathsurround0pt #1\hfil##\hfil$\crcr#2\crcr\sim\crcr}}}
\def\gsim{\wisk{\mathrel{\mathpalette\oversim{>}}}} % > over \sim
\def\lsim{\wisk{\mathrel{\mathpalette\oversim{<}}}} % < over \sim

%%%%%%%%%%%%%%%%%%%%%%%%%%%%%%%%%%%%%%%%%%%%%%%%%%%%%%%%%%%%%%%%%%%%%%%%%%%%%
\newcount\levelone    \levelone=0
\newcount\leveltwo    \leveltwo=0
\newcount\levelthree  \levelthree=0
\newcount\levelfour   \levelfour=0
\def\chaphead{}                             % needed for appendix
\def\secno{\chaphead\the\levelone}
\def\subno{\chaphead\the\levelone.\the\leveltwo}
\def\subsubno{\chaphead\the\levelone.\the\leveltwo.\the\levelthree}
\def\subsubsubno{\chaphead\the\levelone.\the\leveltwo.\the\levelthree
                           .\the\levelfour}
\def\newsec{\advance\levelone by1 \leveltwo=0 \levelthree=0 \levelfour=0}
\def\newsub{\advance\leveltwo by1 \levelthree=0 \levelfour=0}
\def\newsubsub{\advance\levelthree by1 \levelfour=0}
\def\newsubsubsub{\advance\levelfour by1}
%%%%%%%%%%%%%%%%%%%%%%%%%%%%%%%%%%%%%%%%%%%%%%%%%%%%%%%%%%%%%%%%%%%%%%%%%%%%%%
%                                                                            %
%    Definitions of titles of Sections. Tailored for A&A.                    %
%    Remark that we really need unexpanded boldface.  Use expanded for now.  %
%                                                                            %
%%%%%%%%%%%%%%%%%%%%%%%%%%%%%%%%%%%%%%%%%%%%%%%%%%%%%%%%%%%%%%%%%%%%%%%%%%%%%%
\def\absnarrower{\advance\leftskip by \abstractindent}
%         \advance\rightskip by \abstractindent}
\def\titlehang{\hangindent\abstractindent \hangafter 0 \relax}

\newdimen\bottomtol \bottomtol=0.03\vsize
\def\secskip{\par \ifdim\lastskip<\secskipamount \removelastskip \fi
    \vskip 0pt plus \bottomtol \penalty-250
    \vskip 0pt plus -\bottomtol \relax
    \vskip\secskipamount plus3pt minus3pt}
\def\subskip{\par \ifdim\lastskip<\subskipamount \removelastskip \fi
    \vskip 0pt plus 0.5\bottomtol \penalty-150
    \vskip 0pt plus -0.5\bottomtol \relax
    \vskip\subskipamount plus2pt minus2pt}
\long\def\aaabstract#1{\centerline{\null}
   \vskip 1.52cm 
   {\absnarrower \noindent {\bf Summary.} #1 \par}
   \oneskip \oneskip}
\outer\def\unnumberedsectionbegin #1\par {\secskip \noindent {\bf #1}
    \nobreak \vskip 6pt \noindent}
\outer\def\sectionbegin #1\par {\secskip \newsec \noindent {\bf \secno.~#1}
    \nobreak \vskip 6pt \noindent}
\outer\def\subsectionbegin #1\par {\subskip \newsub \noindent {\it \subno.~#1}
    \nobreak \vskip 6pt \noindent}
\outer\def\unnumberedsubsectionbegin #1\par {\subskip \noindent {\it #1}
    \nobreak \vskip 6pt \noindent}
\outer\def\subsubsectionbegin #1\par {\subskip \newsubsub \noindent 
    {\rm \subsubno.~#1}
  \nobreak \vskip 6pt \noindent}
\def\backskip {\vskip -18 pt \relax}
\def\aatitle#1\par {{\null \fourteenpoint 
     \vskip 50pt \baselineskip 18pt \titlehang \noindent \bf #1}
     \ifforcopyeditor \vfil \vfil \fi}
\def\aaauthor#1\par {\vskip 16pt \noindent {\bf #1 \par}
     \ifforcopyeditor \vfil \fi}
\def\aainstitution#1\par {\smallskip \noindent {\rm #1 \par}
     \ifforcopyeditor \vfil \vfil \fi}
\def\keywords#1\par {\oneskip {\ifforcopyeditor \narrower \fi 
  \noindent {\bf Key words: \rm #1}}
  \ifforcopyeditor \vskip 0pt plus 10 fil \relax \eject 
     \else \oneskip \hrule height\ruleht \relax \vskip 15pt \fi
}
%
%%%%%%%%%%%%%%%%%%%%%%%%%%%%%%%%%%%%%%%%%%%%%%%%%%%%%%%%%%%%%%%%%%%%%%%%%%%%%
%                                                                           %
%    Initialization                                                         %
%                                                                           %
%%%%%%%%%%%%%%%%%%%%%%%%%%%%%%%%%%%%%%%%%%%%%%%%%%%%%%%%%%%%%%%%%%%%%%%%%%%%%
\newcount\notenumber
\notenumber=1
\newcount\eqnumber
\eqnumber=1
\newcount\fignumber
\fignumber=1
\newcount\tabnumber
\tabnumber=1
\newbox\abstr
%
%%%%%%%%%%%%%%%%%%%%%%%%%%%%%%%%%%%%%%%%%%%%%%%%%%%%%%%%%%%%%%%%%%%%%%%%%%%%%%
%                                                                            %
%    Equation numbering                                                      %
%    \new macro produces sequentially numbered equations                     %
%    by writing \eqno(\new) at end of displayed equations                    %
%                                                                            %
%%%%%%%%%%%%%%%%%%%%%%%%%%%%%%%%%%%%%%%%%%%%%%%%%%%%%%%%%%%%%%%%%%%%%%%%%%%%%%
\def\new{{\rm\chaphead\the\eqnumber}\global\advance\eqnumber by 1}
%%%%%%%%%%%%%%%%%%%%%%%%%%%%%%%%%%%%%%%%%%%%%%%%%%%%%%%%%%%%%%%%%%%%%%%%%%%%%%
%                                                                            %
%    to refer to an equation which is 5 equations back,                      %
%    write "equation (\eqref5)"                                              %
%                                                                            %
%%%%%%%%%%%%%%%%%%%%%%%%%%%%%%%%%%%%%%%%%%%%%%%%%%%%%%%%%%%%%%%%%%%%%%%%%%%%%%
\def\eqref#1{\advance\eqnumber by -#1 \chaphead\the\eqnumber
           \advance\eqnumber by #1 }
\def\?{\eqref{1}}
%%%%%%%%%%%%%%%%%%%%%%%%%%%%%%%%%%%%%%%%%%%%%%%%%%%%%%%%%%%%%%%%%%%%%%%%%%%%%%
%                                                                            %
%    \last macro is like \new except counter is not advanced. Useful for     %
%    equations which are in parts a and b.                                   %
%                                                                            %
%%%%%%%%%%%%%%%%%%%%%%%%%%%%%%%%%%%%%%%%%%%%%%%%%%%%%%%%%%%%%%%%%%%%%%%%%%%%%%
\def\last{\advance\eqnumber by -1 {\rm\chaphead\the\eqnumber}\advance
     \eqnumber by 1}
%%%%%%%%%%%%%%%%%%%%%%%%%%%%%%%%%%%%%%%%%%%%%%%%%%%%%%%%%%%%%%%%%%%%%%%%%%%%%%
%                                                                            %
%    to name an equation, place command "\eqnam{\Poisson}" before equation,  %
%    and thereafter "equation(\Poisson)" will generate the proper equation   %
%    number.                                                                 %
%                                                                            %
%%%%%%%%%%%%%%%%%%%%%%%%%%%%%%%%%%%%%%%%%%%%%%%%%%%%%%%%%%%%%%%%%%%%%%%%%%%%%%
\def\eqnam#1{\xdef#1{\chaphead\the\eqnumber}}
%%%%%%%%%%%%%%%%%%%%%%%%%%%%%%%%%%%%%%%%%%%%%%%%%%%%%%%%%%%%%%%%%%%%%%%%%%%%%%
%                                                                            %
%    For the Appendix                                                        %
%                                                                            %
%%%%%%%%%%%%%%%%%%%%%%%%%%%%%%%%%%%%%%%%%%%%%%%%%%%%%%%%%%%%%%%%%%%%%%%%%%%%%%
%                                                                            %
\def\appendixbegin#1 #2{\eqnumber=1 \def\chaphead{{#1}}
    \levelone=0\leveltwo=0\levelthree=0\levelfour=0\eqnumber=1\fignumber=1 
    \vskip\subskipamount\noindent{\ninepoint\bf Appendix #1\ \ \ #2}
    \vskip\subskipamount\noindent}
\def\noappendixbegin#1 #2{\eqnumber=1 \def\chaphead{{#1} }
    \levelone=0\leveltwo=0\levelthree=0\levelfour=0\eqnumber=1\fignumber=1 
    \vskip\subskipamount\noindent{}
    \vskip\subskipamount\noindent}
%%%%%%%%%%%%%%%%%%%%%%%%%%%%%%%%%%%%%%%%%%%%%%%%%%%%%%%%%%%%%%%%%%%%%%%%%%%%%%
%                                                                            %
%    figure numbering                                                        %
%    \nfig macro assigns number to a figure                                  %
%                                                                            %
%%%%%%%%%%%%%%%%%%%%%%%%%%%%%%%%%%%%%%%%%%%%%%%%%%%%%%%%%%%%%%%%%%%%%%%%%%%%%%
\def\nfig{\chaphead\the\fignumber\global\advance\fignumber by 1}
\def\ntab{\chaphead\the\tabnumber\global\advance\tabnumber by 1}
%%%%%%%%%%%%%%%%%%%%%%%%%%%%%%%%%%%%%%%%%%%%%%%%%%%%%%%%%%%%%%%%%%%%%%%%%%%%%%
%                                                                            %
%    \nfiga permits a,b,c etc. to be added to figure number                  %
%                                                                            %
%%%%%%%%%%%%%%%%%%%%%%%%%%%%%%%%%%%%%%%%%%%%%%%%%%%%%%%%%%%%%%%%%%%%%%%%%%%%%%
\def\nfiga#1{\chaphead\the\fignumber{#1}\global\advance\fignumber by 1}
\def\rfig#1{\advance\fignumber by -#1 \chaphead\the\fignumber
            \advance\fignumber by #1}
\def\fignam#1{\xdef#1{\chaphead\the\fignumber}}
\def\tabnam#1{\xdef#1{\chaphead\the\tabnumber}}
%
%    References (in AA style)
%

\def\spirnir#1 {, {1996, In: {Minniti, Rix (eds.)
      Spiral galaxies in the NIR}. Heidelberg, p. #1 }}

\def\cengal#1{, {1989, In: {Morris M.(ed.) Proc. IAU Symp. 136,
     The Centre of the Galaxy.} Kluwer, p. #1}}

\def\lodm#1.{, 1986, In: {Israel F.P. (ed.) Light on Dark Matter.}
  Reidel, Dordrecht, p. #1}
\def\lsse#1.{, 1987, In: {Kwok S., Pottasch S.R. (eds.) Late stages of
   stellar evolution.} Reidel, Dordrecht, p. #1}
\def\galaxy#1.{, 1987, In: {Gilmore G., Carswell B.(eds.) Galaxy.}
   Reidel, Dordrecht, p. #1}
\def\torc#1.{, 1988, In: {Fich M.(ed.) Mass of the Galaxy.}
  Toronto University Press, p. #1}
\def\planneb#1.{, 1989, In: {Torres--Peimbert (ed.) Planetary Nebulae.},
  Reidel, Dordrecht, p. #1}
\def\adass#1 {, {1995, In: {Shaw R.A., Payne H.E., Hayes J.J.E. (eds.) PASPC 77,
   Astronomical Data Analysis Software and Systems IV, } p. #1}}
\def\plarin#1.{, 1984, In: {Greenberg R., Brahic A. (eds.) Planetary Rings.},
  Tucson, p. #1}

\def\seng#1 {, {1981, In: { 
      The structure and evolution of normal galaxies}. Cambridge, p. #1 }}

\def\varmic#1 {, {In: {Ferlet, Maillard, Raban (eds.) 
     Variable stars and astrophysical returns of microlensing surveys}. 
     Ed. Fronti\`eres, p. #1 }}

\def\solve#1 {, {1996, In: {Blitz L., Teuben P.(eds.) Proc.
     IAU Symp. 169, Unsolved problems of the Milky Way}.
     Reidel, Dordrecht, p. #1 }}

\def\mosgn#1 {, {1988, In: {Bianchi, Gilmozzi (eds.)
  Mass outflow from stars and Galactic Nuclei.} p. #1 }}
\def\pprg#1 {, {1981, In: {Iben, Renzini (eds.) Physical Processes in
      Red Giants. } p. #1 }}
\def\cbdmw#1 {, {1992, In: {Blitz (ed.) The Center, Bulge and Disk of the
      Milky Way.} Kluwer, Dordrecht, p. #1 }}
\def\planeb#1 {, {1993, In: {Weinberger R., Acker A.(eds.)
      Proc. IAU Symp. 155,
      Planetary Nebulae.} Reidel, Dordrecht, p. #1 }}
\def\mpneb#1 {, {1990, In: {Mennessier M.O., Omont A. (eds.)
      From Miras to Planetary Nebulae. Yvette Cedex: \'Editions
      Fronti\`eres, } p. {#1}\ }}
\def\gents#1 {, {1993, In: {Dejonghe H., Habing H.J. (eds.) Proc.
     IAU Symp. 153, Galactic Bulges}. Reidel, Dordrecht, p. #1 }}
\def\bargal#1 {, {1996, In: {Buta, Crocker, Elmegreen (eds.)
      PASPC 91, Barred Galaxies, } p. #1\ }}
\def\mopste#1 {, {1994, In: Jorgensen U.G. (ed.) Proc. IAU Coll. 146,
     Molecular Opacities in the Stellar Environment. Springer-Verlag, p. #1}}
\def\galstr#1 {, {1965, In: {Blaauw A., Schmidt M. (eds.)
     Galactic Structure}. Chicago, p. #1 }}

\def\physr#1 {, {\it Physics Report}{\bf#1},\ }
\def\iauc#1 #2 {, {IAU Circ.\ }{#1, #2}\ }
\def\nature#1 #2 {, {Nat \ }{#1, #2}\ }
\def\science#1 #2 {, {Sci \ }{#1, #2}\ }
\def\aa#1 #2 {, {A\&A}{ #1, #2} }
\def\aal#1 #2 {, {A\&A}{ #1, L#2}\ }
\def\aas#1 #2 {, {A\&AS}{ #1, #2} }
\def\aj#1 #2 {, {AJ\ }{#1, #2}\ }
\def\apj#1 #2 {, {ApJ\ }{#1, #2}\ }
\def\apjl#1 #2 {, {ApJ\ }{#1, L#2 }\ }
\def\apjs#1 #2 {, {ApJS\ }{#1, #2}\ }
\def\araa#1 #2 {, {ARA\&A\ }{#1, #2}\ }
\def\araapr{, {ARA\&A\ }{in preparation}\ }
\def\mnras#1 #2 {, {MNRAS\ }{#1, #2}\ }
\def\mnrasprep {, {MNRAS\ }{in preparation}\ }
\def\rpphys#1 #2 {, {Rep. Prog. Phys.\ }{#1, #2}\ }

\def\pasp#1 #2{, {PASP \ }{#1, #2}\ }
\def\qjras#1 {, {QJRAS \ }{#1}\ }
\def\aus#1 #2{, {Aust.~J. Phys.\ }{#1, #2}\ }
\def\actaa#1 {, {Acta Astron.\ }{#1}\ }

\def\refBinGerDep{Binney \& Gerhard 1995}
%Binney and Gerhard 1995\mnrasprep\ 

%% On the derpojection of the Galactic Bulge, see also
%% the  Photometric structure of the inner Galaxy (B,G + Spergel)

\def\refvdV1989{van der Veen 1989}
%van der Veen W., 1989\aa 210 127 

\def\refCOBEDwek{Dwek \etal\ 1995}
%Dwek \etal\ 1995\apj 445 716 

\def\refvdVH1990{van der Veen \& Habing 1990}
%van der Veen W., Habing H.J., 1990\aa 231 404\ 

%%\def\refZSR1994{Zhao H.S., Spergel D.N., Rich M., 1994\aj 108 2154\ }

\def\refBlom {Blommaert 1992}
%Blommaert 1992

\def\refHarmrev{Habing 1996}
%Habing H.J., 1996\araa 7 97

\def\refHarmgent{Habing 1993}
%Habing H.J. \gents 57

\def\refWilBar{Wilson \& Barrett 1968}
%Wilson, W.J., Barett, A.H., 1968\science 161 778

\def\refRenz{Renzini 1981}
%Renzini \pprg 431

\def\refPAWMF{Whitelock \& Feast 1993}
%Whitelock P.A., Feast M. \planeb 251

\def\refVasW{Vassiliadis \& Wood 1993}
%Vassiliadis E., Wood P.R., 1993\apj 413 641

\def\refEGS{Elitzur \etal\ 1976}
%Elitzur M., Goldreich P., Scoville N., 1976\apj 205 384

\def\refOlof{Olofsson 1994}
% Olofsson H. \mopste 113

\def\refCohrp{Cohen 1989}
%% Cohen R.J., 1989\rpphys 52 881

\def\refQP{Dejonghe 1989}
%%Dejonghe, H., 1989\apj 343 113

\def\refBT{Binney \& Tremaine 1987}
%%Binney \& Tremaine, Galactic Dynamics}

\def\refBS1{ Blitz \& Spergel 1991}
%%Blitz \& Spergel 1991\apj 379 631

\def\refABC{Aaronson etal. 1989,1990}
%%13:Aaronson, Blanco, Cook, Schechter\1989\apjs 70 637
%%16:Aaronson, Blanco, Cook, Olszewksi, Schechter,1990\apjs 73 841

\def\refAVK{Cardelli etal. 1988}
%%Cardelli J.A., Clayton G.C., Mathis J.S.  1988\apj 345 245

\def\refAVform{Milne \& Aller 1980}
%%Milne D.K., Aller L.H., 1980\aj 85 17

\def\mum{\wisk{\mu}m}
\def\kms{\wisk{\,\rm km\,s^{-1}\,}}                    % km s-1
\def\decdeg#1.#2 {\wisk{#1^{\,\rm o}\bck.\,#2}\ }
\def\losa{line--of--sight}
\def\losn{line of sight}
\def\degr{\wisk{^{\circ}}}                                % degrees symbol
\def\deg{\wisk{^{\circ}}}                                % degrees symbol
\def\decsec#1.#2 {\wisk{#1^{\prime\prime}\hskip-0.42em.\hskip0.10em#2}\ }
\def\kmsr{\wisk{\,\rm km\,s^{-1}\,kpc^{-1}}}
\def\pspeed{\wisk{ \Omega_{\rm p}}}

\def\etal{{et al.$\,$}}
\def\cse{circum--stellar envelope}
\def\cses{circum--stellar envelopes}
\def\df{distribution function}
\def\dfs{distribution functions}
\def\lvd{longitude--velocity diagram}
\def\lvds{longitude--velocity diagrams}

\def\rsun{\wisk{\rm \,R_\odot}}
\def\vsun{\wisk{\rm \,V_\odot}}
\def\msun{\wisk{\rm \,M_\odot}}
\def\msol{\wisk{\rm \,M_\odot}}

\def\vexp{\wisk{\,V_{\rm exp}}}
\def\vlsr{\wisk{\,V_{\rm LSR}}}
\def\gbu{galactic Bulge}
\def\gd{galactic Disk}
\def\gba{galactic Bar}
\def\gc{galactic Centre}

\def\dpk{double--peaked}
\def\spk{single--peaked}

\def\msyr{\wisk{\,\rm M_\odot\,yr^{-1}}}